\documentclass{emulateapj}

\shorttitle{PTF09dav}
\shortauthors{Sullivan et al.}

\newcommand{\synapps}{\texttt{SYNAPPS}}

\newcommand{\synow}{\texttt{SYNOW}}

\newcommand{\zhel}{\ensuremath{z_{\mathrm{hel}}}}
\newcommand{\zcmb}{\ensuremath{z_{\mathrm{cmb}}}}
\newcommand{\dmb}{\ensuremath{\Delta m_{15}(B)}}
\newcommand{\nickel}{\ensuremath{^{56}\mathrm{Ni}}}
\newcommand{\mni}{\ensuremath{M_{\mathrm{Ni}}}}
\newcommand{\lbol}{\ensuremath{L_{\mathrm{bol}}}}
\newcommand{\radlum}{\ensuremath{\dot{S}}}
\newcommand{\msun}{\ensuremath{M_{\odot}}}

\begin{document}


\title{The Subluminous and Peculiar Type Ia Supernova PTF\,09dav}


\author {
  M.~Sullivan\altaffilmark{1},
  M.~M.~Kasliwal\altaffilmark{2},
  P.~E.~Nugent\altaffilmark{3},
  D.~A.~Howell\altaffilmark{5,6},
  R.~C.~Thomas\altaffilmark{3},
  E.~O.~Ofek\altaffilmark{1},
  I.~Arcavi\altaffilmark{4},
  S.~Blake\altaffilmark{1},
  J.~Cooke\altaffilmark{2},
  A.~Gal-Yam\altaffilmark{4},
  I.~M.~Hook\altaffilmark{7},
  P.~Mazzali\altaffilmark{8,9,10},
  P.~Podsiadlowski\altaffilmark{1},
  R.~Quimby\altaffilmark{2},
  L.~Bildsten\altaffilmark{6,11},
  J.~S.~Bloom\altaffilmark{12},
  S.~B.~Cenko\altaffilmark{12},
  S.~R.~Kulkarni\altaffilmark{2},
  N.~Law\altaffilmark{13},
  D.~Poznanski\altaffilmark{3,12}
}
\altaffiltext{1}{Department of Physics (Astrophysics), University of Oxford, Keble Road, Oxford OX1 3RH, UK}
\altaffiltext{2}{Cahill Center for Astrophysics, California Institute of Technology, Pasadena, CA, 91125, USA}
\altaffiltext{3}{Computational Cosmology Center, Lawrence Berkeley National Laboratory, 1 Cyclotron Rd., Berkeley CA 94720, USA}
\altaffiltext{4}{Department of Particle Physics and Astrophysics, Faculty of Physics, The Weizmann Institute of Science, Rehovot 76100, Israel}
\altaffiltext{5}{Las Cumbres Observatory Global Telescope Network, 6740 Cortona Dr, Suite 102, Goleta, CA 93117}
\altaffiltext{6}{Department of Physics, University of California Santa Barbara, Santa Barbara, CA 93106, USA}
\altaffiltext{7}{INAF-Osservatorio di Roma, via Frascati 33, I-00040 Monteporzio Catone (Roma), Italy}
\altaffiltext{8}{Max-Planck-Institut für Astrophysik, Karl-Schwarzschild-Str. 1, 85741 Garching, Germany}
\altaffiltext{9}{Scuola Normale Superiore, Piazza dei Cavalieri 7, 56126 Pisa, Italy}
\altaffiltext{10}{National Institute for Astrophysics-OAPd, Vicolo dell'Osservatorio 5, 35122 Padova, Italy}
\altaffiltext{11}{Kavli Institute for Theoretical Physics, University of California Santa Barbara, Santa Barbara, CA 93106, USA}
\altaffiltext{12}{Department of Astronomy, University of California, Berkeley, CA 94720-3411, USA}
\altaffiltext{13}{Dunlap Institute for Astronomy and Astrophysics, University of
Toronto, 50 St. George Street, Toronto M5S 3H4, Ontario, Canada}

\email{sullivan@astro.ox.ac.uk}

\begin{abstract}

  PTF\,09dav is a peculiar subluminous type Ia supernova (SN)
  discovered by the Palomar Transient Factory (PTF).
  Spectroscopically, it appears superficially similar to the class of
  subluminous SN1991bg-like SNe, but it has several unusual features
  which make it stand out from this population. Its peak luminosity is
  fainter than any previously discovered SN1991bg-like SN Ia
  ($M_B\sim-15.5$), but without the unusually red optical colors
  expected if the faint luminosity were due to extinction. The
  photospheric optical spectra have very unusual strong lines of
  \ion{Sc}{2} and \ion{Mg}{1}, with possible \ion{Sr}{2}, together
  with stronger than average \ion{Ti}{2} and low velocities of
  $\sim6000$\,km\,s$^{-1}$.  The host galaxy of PTF09dav is ambiguous.
  The SN lies either on the extreme outskirts ($\sim41$\,kpc) of a
  spiral galaxy, or in an very faint ($M_R\geq-12.8$) dwarf galaxy,
  unlike other 1991bg-like SNe which are invariably associated with
  massive, old stellar populations. PTF\,09dav is also an outlier on
  the light-curve-width--luminosity and color--luminosity relations
  derived for other sub-luminous SNe Ia.  The inferred \nickel\ mass
  is small ($0.019\pm0.003\msun$), as is the estimated ejecta mass of
  $0.36\msun$.  Taken together, these properties make PTF\,09dav a
  remarkable event.  We discuss various physical models that could
  explain PTF\,09dav. Helium shell detonation or deflagration on the
  surface of a CO white-dwarf can explain some of the features of
  PTF\,09dav, including the presence of Sc and the low photospheric
  velocities, but the observed Si and Mg are not predicted to be very
  abundant in these models.  We conclude that no single model is
  currently capable of explaining all of the observed signatures of
  PTF\,09dav.

\end{abstract}


\keywords{supernovae: general --- supernovae: individual (PTF09dav)}

\section{Introduction}
\label{sec:introduction}

Type Ia supernovae (SNe Ia), thermonuclear explosions of accreting
carbon-oxygen white dwarf stars, form a fairly uniform and homogeneous
class of events. This has inspired their application as cosmological
standardizable candles, tracing the cosmic expansion history to high
redshift. However, the class of ``sub-luminous'' SN Ia events show
quite different photometric and spectroscopic properties than those
mainstream events used for cosmological studies, with potentially
different progenitor models.  Though the discovery of these rare,
faint SNe Ia has historically been challenging, the archetypal
sub-luminous SN1991bg \citep{1992AJ....104.1543F,1993AJ....105..301L}
has been complemented with the discovery of many similar events
\citep[e.g., see compilations
of][]{1993AJ....106.2383B,2004ApJ...613.1120G,2008MNRAS.385...75T,2010arXiv1011.4531G}.
Modern SN searches are revealing ever more extreme examples of faint
SNe of all types, the nature of which, in some cases, remains
uncertain
\citep[e.g.,][]{2003PASP..115..453L,2008ApJ...683L..29K,2009Natur.459..674V,2009AJ....138..376F,2010Natur.465..322P,2010ApJ...708L..61F,2010ApJ...720..704M,2010ApJ...723L..98K}.

The key characteristics of the 1991bg-like sub-luminous SN Ia
population are a faint peak absolute magnitude ($\sim2$ magnitudes
fainter than a normal SN Ia) with a red optical color, a light curve
that rises and falls $\sim40$\% faster than a normal SN Ia, a strong
trend of being located in old, E/S0 galaxies with large stellar masses
\citep{2001ApJ...554L.193H,2009ApJ...707.1449N}, and `cool' optical
maximum light spectra dominated by strong \ion{Ti}{2} absorption
\citep{1992AJ....104.1543F,1997MNRAS.284..151M,2006PASP..118..560B}
with somewhat lower expansion velocities than in normal SNe Ia
\citep{2005ApJ...623.1011B}.  The class of SNe Ia similar to SN2002cx
\citep{2003PASP..115..453L} appear to form a further, separate
sub-class of faint events, with a host galaxy distribution favoring
later morphological types compared to that of SN1991bg-like events
\citep{2009AJ....138..376F,2009Natur.459..674V}. Some events in the
sub-class, such as SN2008ha, are fainter still than 1991bg-like SNe
\citep{2009AJ....138..376F,2010ApJ...720..704M}, with even lower
ejecta velocities (and kinetic energies). Recent observations also
demonstrate the existence of further faint or fast types of SN events,
where the formal SN type classification may provide mis-leading
information about the physical nature of the explosion
\citep{2010Sci...327...58P,2010Natur.465..322P,2010arXiv1008.2754P}.

Various physical models and scenarios have been proposed to explain
sub-luminous SN1991bg-like events. Whether they form a distinct
physical group from normal SNe Ia, with different progenitors and
explosion models, or whether they lie at the extreme end of a
continuous distribution, but with lower \nickel\ masses and hence
temperatures, is unclear. No conclusive evidence for any one
progenitor model has so far been found. The nature of the progenitors
of SN2002cx-like events is even less clear, with possible models
ranging from the direct collapse of a massive star to a black hole
\citep[a ``fallback'' SN;][]{2009Natur.459..674V,2010ApJ...719.1445M},
to the pure deflagration of a white dwarf
\citep[e.g.][]{2004PASP..116..903B,2007PASP..119..360P}.

In this paper we describe the supernova PTF\,09dav. This SN was
discovered as part of the Palomar Transient Factory
\citep[PTF\footnote{\url{http://www.astro.caltech.edu/ptf/}};][]{2009PASP..121.1334R,2009PASP..121.1395L},
a five year project surveying the optically transient sky. PTF is
designed to sample a large fraction of the optical transient
population, including both new types of events, as well as
statistically complete samples of known transient types.  In
particular, the high-cadence, survey depth, and wide-area make PTF
ideal for the study of fast, subluminous SN events
\citep[e.g.][]{2010ApJ...723L..98K}. PTF\,09dav is one such event -- a
very subluminous SN Ia with peculiar (and so far unique) spectral
properties. In $\S$~\ref{sec:observations} we introduce PTF\,09dav
together with its basic photometric, spectral, and host galaxy data.
We analyze the light curve in $\S$~\ref{sec:light-curve-analysis},
deriving estimates of its peak luminosity and related parameters.
$\S$~\ref{sec:spectral-analysis} details our spectral analysis,
including the identification of the principal elements. We discuss
various physical interpretations of our data in
$\S$~\ref{sec:discussion}. In a companion paper, we discuss the
late-time light curve and nebular spectroscopic data for this event
(Kasliwal et al., in prep.). Where relevant we assume
$H_0$=70~km~s$^{-1}$~Mpc$^{-1}$.

\section{Observations}
\label{sec:observations}

\begin{figure}
\centering
\includegraphics[width=0.495\textwidth]{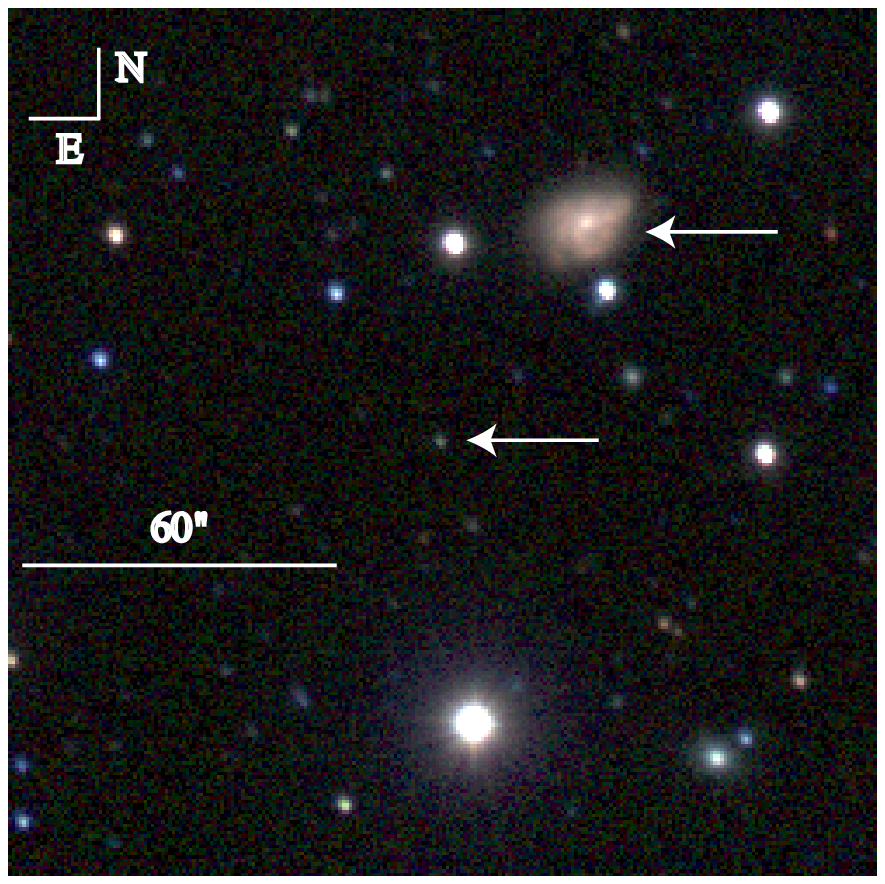}
\caption{A false color image constructed from P60 $g$, $r$ and $i$
  data showing the location of PTF\,09dav in relation to the nearby
  ``host galaxy''. The upper arrow indicates the putative host, and
  the lower arrow the SN. The separation is 56.8\arcsec, which at
  $\zcmb=0.0359$ corresponds to 40.6\,kpc. The host galaxy has an
  apparently disturbed morphology, and a spectrum shows it to be
  dominated by nebular emission lines indicative of ongoing star
  formation. No host is detected at the position of PTF\,09dav to
  $m_{\mathrm{RG610}}=23.2$
  ($M_{\mathrm{RG610}}=-12.8$).\label{fig:host}}
\end{figure}

PTF\,09dav was discovered as a transient event at $R\sim19.5$ by the
Palomar Transient Factory (PTF) on 2009 August 11.3 (all dates in this
paper are UTC), located at $\alpha=22^{\mathrm{h}}46^{\mathrm{m}}55\fs
15$, $\delta=+21\arcdeg37\arcmin34\farcs 1$ (J2000;
Fig.~\ref{fig:host}).  The event was isolated with no apparent host
galaxy in the PTF reference images.  A spectrum was taken with the
William Herschel Telescope (WHT) using the Intermediate dispersion
Spectrograph and Imaging System (ISIS) on 2009 August 14.0 as part of
the PTF SN Ia key project. The R316R (red arm) and R300B (blue arm)
gratings were used, together with the 5300 dichroic, giving a
wavelength coverage of $\sim$3400--8100\AA. Spectral comparisons were
performed at the telescope using the \texttt{SUPERFIT} spectral
matching code \citep[e.g.,][]{2005ApJ...634.1190H}, and the spectrum
was initially recognized as being superficially similar to the
subluminous SN Ia SN1991bg \citep{1992AJ....104.1543F}, with deep
\ion{Ti}{2}, \ion{Si}{2}, and \ion{O}{1} absorption, located at a
redshift of $z\sim0.04$ (Fig.~\ref{fig:obsspectra}).  Given the faint
discovery magnitude and relatively low redshift, PTF\,09dav was
recognized as a potentially rare and interesting sub-luminous SN
event, and was triggered for further study, which we summarize in the
remainder of this section.

\subsection{Spectra}
\label{sec:spectra}

Following the initial WHT classification spectrum, two further spectra
were taken during the photospheric phase of the SN. The first was
taken with the Palomar Hale 200-in telescope (P200) plus Double Beam
Spectrograph \citep[DBSP;][]{1982PASP...94..586O} on 2009 August 20.5.
The 600l/4000\AA\ (blue) and 158l/7500\AA\ (red) gratings, and D55
dichroic, were used, giving a wavelength coverage of
$\sim$3500--9500\AA. The second spectrum was taken using the Keck-I
telescope and the Low-Resolution Imaging Spectrometer
\citep[LRIS;][]{1995PASP..107..375O} on 2009 August 25.5 -- a
wavelength coverage of $\sim$3400--10000\AA\ was achieved using the
400l/3400\AA\ grism (blue), 400l/8500\AA\ grating (red), and the 560
dichroic. A final spectrum taken during the nebular phase of the SN
was taken on 2009 November 11.4, and will be presented in a companion
paper (Kasliwal et al. in prep.).

The three photospheric spectra were reduced using the same
custom-written pipeline based on standard procedures in \texttt{IRAF}
and \texttt{IDL} broadly following the reduction procedures outlined
in \citet{2008ApJ...674...51E}, including flux calibration and
telluric feature removal.  All spectra were observed at the
parallactic angle in photometric conditions.  ``Error'' spectra are
derived from a knowledge of the CCD properties and Poisson statistics,
and are tracked throughout the reduction procedure. The spectra are
also rebinned (in a weighted way) to a common dispersion, and joined
across the dichroic.  The three spectra are shown in
Figs.~\ref{fig:obsspectra} and \ref{fig:obsspectra2}.

\subsection{Host galaxy}
\label{sec:host-galaxy}

The host galaxy of PTF\,09dav is somewhat ambiguous
(Fig.~\ref{fig:host}). PTF\,09dav lies 56.8\arcsec\ (projected) SE of
a potential host galaxy ($r\simeq17.6$;
$\alpha=22^{\mathrm{h}}46^{\mathrm{m}}52\fs 9$,
$\delta=+21\arcdeg38\arcmin21\farcs 6$, J2000). A Lick spectrum taken
on 2009 October 25th shows this galaxy to have strong nebular emission
lines (H$\alpha$, H$\beta$, \ion{O}{3}, \ion{O}{2}) and be located at
a heliocentric redshift $\zhel=0.0371\pm0.0002$ (well within the
Hubble flow), consistent with the estimated SN redshift. At the
position of the SN, $\zhel=0.0371$ corresponds to $z=0.0359$ in the
CMB rest-frame (\zcmb), or a distance modulus of $35.99\pm0.06$ (the
error assumes a host galaxy peculiar velocity uncertainty of
$300\,\mathrm{km}\,\mathrm{s}^{-1}$). This would give a projected
physical separation between PTF\,09dav and the center of this galaxy
of 40.6\,kpc.

There is no evidence for a host galaxy at the position of the SN.
Using 129
``DeepSky''\footnote{\url{http://supernova.lbl.gov/$\sim$nugent/deepsky.html}}
\citep{2009AAS...21346910N} images taken with the Palomar-QUEST survey
over of the field of PTF\,09dav, we have constructed a deep RG610 image
(a longpass $r+i+z$ filter).  Using a 4\arcsec\ diameter aperture at
the position of the SN, we detect no host galaxy with a 3-$\sigma$
limiting magnitude of $m=23.2$ in the RG610, or $M=-12.8$ at
$z=0.0359$. Such a faint absolute magnitude would be lower than any
detected host galaxy of a core collapse SN found by PTF
\citep{2010ApJ...721..777A}. Deeper Keck imaging at the position of
PTF\,09dav is underway (Kasliwal et al., in prep.).

\subsection{Light curve data}
\label{sec:light-curve}

\begin{figure}
\includegraphics[width=0.495\textwidth]{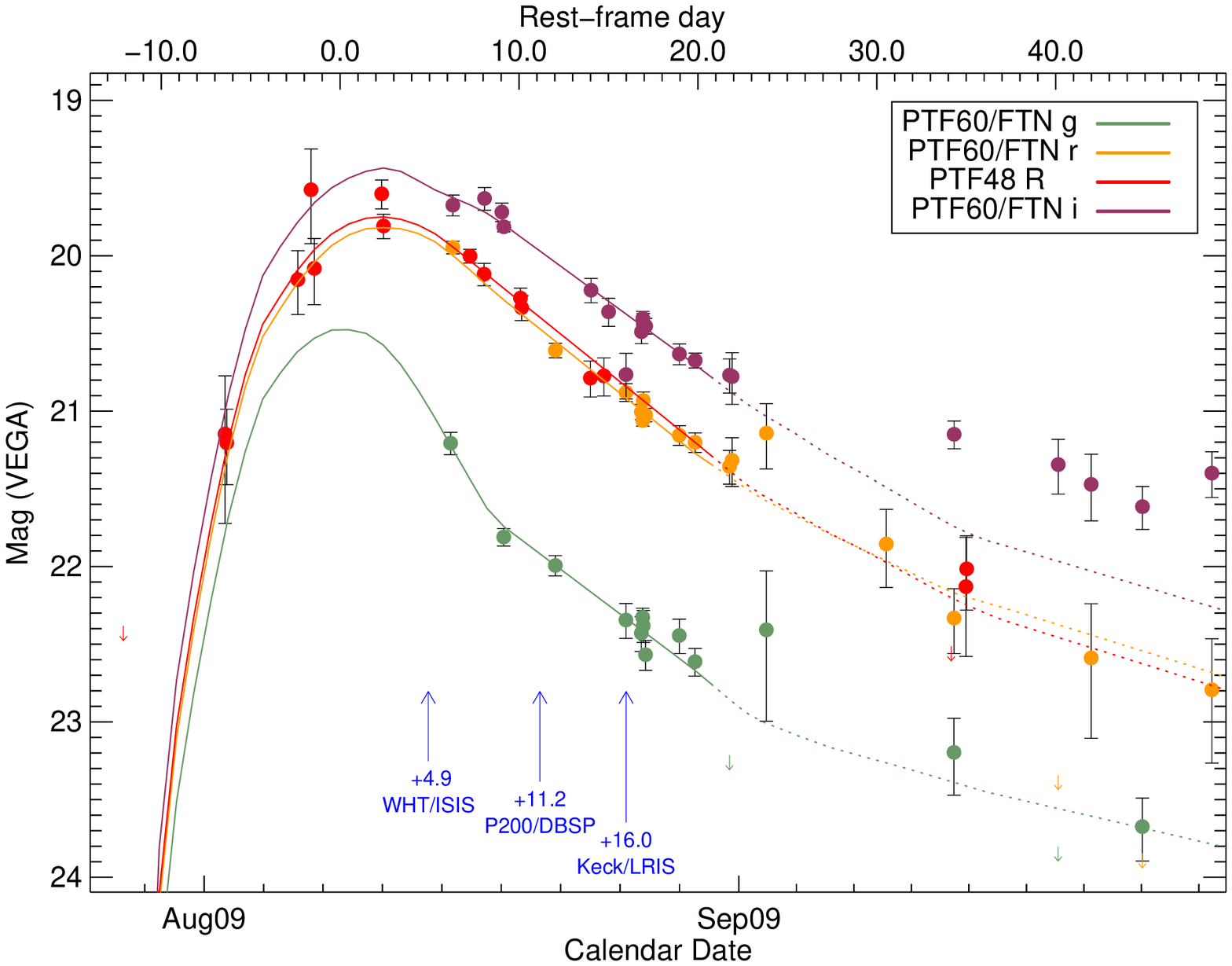}
\includegraphics[width=0.495\textwidth]{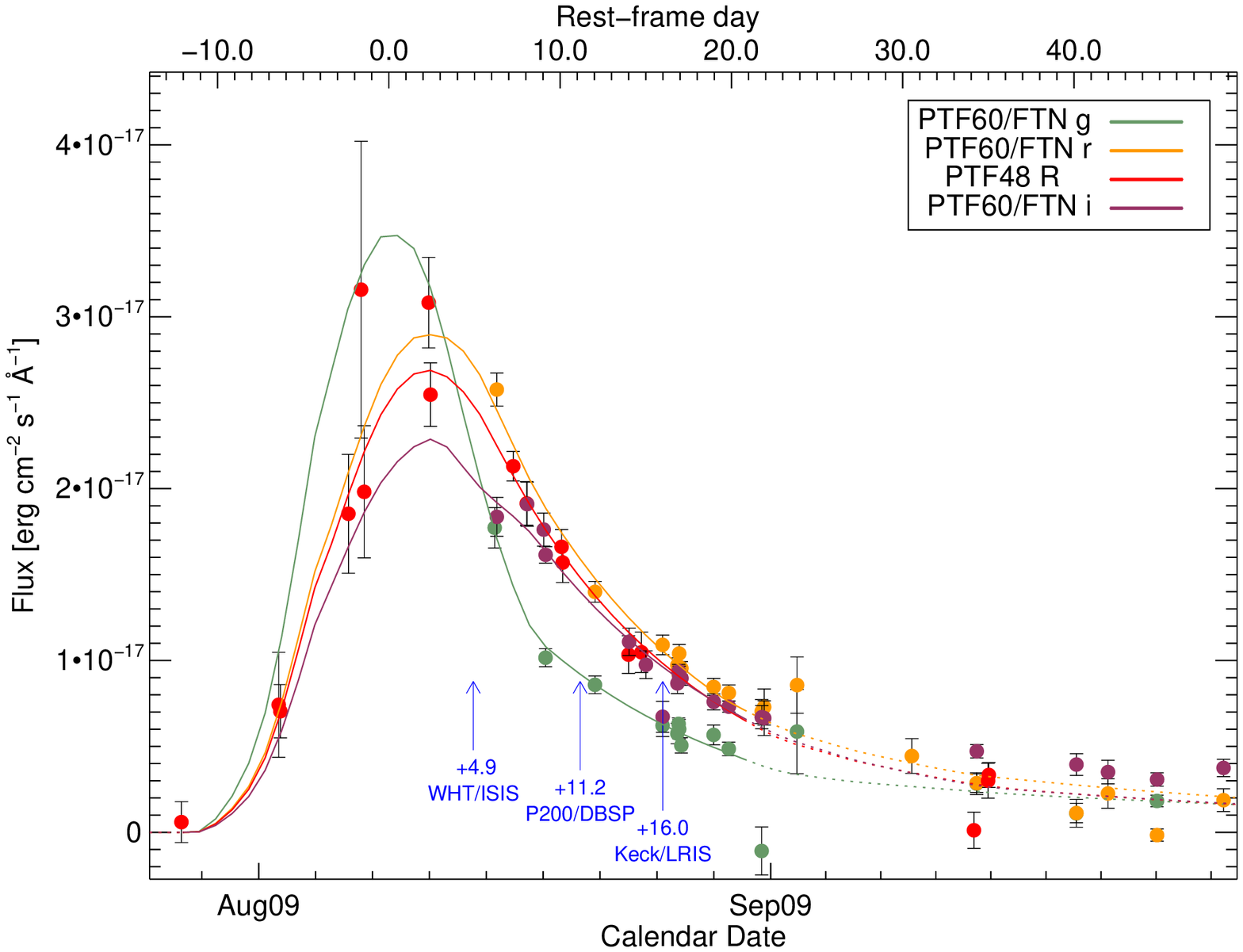}
\caption{The observed light curve in apparent magnitude space (left)
  and flux space (right) for PTF\,09dav in P48 $R$ and P60/FTN $g$,
  $r$, $i$.  For clarity the FTN and P60 data are denoted by the same
  symbol (but the correct filter responses are used in the light curve
  fit). The overlaid lines are the best-fitting SiFTO light curve
  using a sub-luminous SN Ia spectral template, with an arbitrary
  flux-scaling in each filter, but a common ``stretch'' factor and
  time of maximum light. Only datapoints up to +20 days are used in
  the fit; the template lines are dotted after this phase and show the
  predicted evolution at later phases according to the template.
  Maximum light occurs at an MJD of 55051.9. The arrows mark the
  epochs (relative to maximum light in the rest-frame $B$-band) of the
  three photospheric spectra in Fig.~\ref{fig:spectralfits}. The
  reduced $\chi^2$ of the fit is 1.02, and in $gRr$ the light curve
  decay is completely consistent with that of a stretched version of
  SN1991bg-like SN Ia.  Note that at phases later than +30 days, there
  is excess $i$ flux compared to the simple template
  prediction.\label{fig:lightcurve}}
\end{figure}

Photometric monitoring in $gri$ filters commenced using the robotic
Palomar 60in (P60) telescope \citep{2006PASP..118.1396C} on 2009
August 15.3 and on the robotic Faulkes Telescope North (FTN) on 2009
August 18.4, complementing the rolling PTF search on the Samuel Oschin
48in telescope (P48) in $R$-band.  The resulting multi-color light
curve of PTF\,09dav is shown in Fig.~\ref{fig:lightcurve}. As
PTF\,09dav has no detected host galaxy at the position of the SN, no
subtraction of a reference image is required, simplifying the light
curve measurement.  We measure the SN photometry using a
point-spread-function (PSF) fitting method. In each image frame, the
PSF is determined from nearby field stars, and this average PSF is
then fit at the position of the SN event weighting each pixel
according to Poisson statistics, yielding a SN flux and flux error.

We perform a flux calibration to the Sloan Digital Sky Survey (SDSS)
photometric system, close to the AB system
\citep{1983ApJ...266..713O}, using P48 and P60 observations of SDSS
fields made on the same night as observations of PTF\,09dav.  We use
an average calibration from photometric nights in August and September
2009, which provides calibrated tertiary standards in the field of
PTF\,09dav.  This calibration is then transferred to other epochs, and
to the FTN observations, using aperture photometry of these tertiary
standards to determine an relative flux-scaling factor for each epoch,
typically measured to $<$1\%. In the P60/FTN $g$, $r$ and $i$ filters
the color terms to the SDSS system are very small.  The P48 color term
is larger, and we include color, extinction, and color--extinction
terms.  The r.m.s.  of the color term fits is 0.02-0.03\,mag (with a
color term in $r-i$ of $\sim0.22$\,mag), although we present
magnitudes in the natural P48 system and do not apply this color term
to our photometry.  Finally, we fit the same PSF used for the SN flux
measurement to the calibrating field stars, and apply a correction to
go from the aperture in which the flux calibration is determined, to
the PSF fit in which the SN is measured. 

Although we could perform our entire analysis using magnitudes defined
in the AB system, rest-frame SN Ia properties in the literature are
typically given in the Vega system. For ease of comparison to previous
work, we convert our calibrated light curves into this system when
performing the light curve analysis, and all quoted magnitudes are
given in the Vega system. The offsets from Vega ($m_{\mathrm{vega}}$)
to AB ($m_{\mathrm{AB}}$) magnitudes are -0.12, 0.13, 0.19, and 0.35
for the $g$, $r$, $R$, and $i$ filters respectively
($m_{\mathrm{vega}}=m_{\mathrm{AB}}-\mathrm{offset}$).  Our light
curve data can be found in Table~\ref{tab:photometry}, presented in
counts rather than magnitudes to preserve information on
non-detections.

\section{Light curve analysis}
\label{sec:light-curve-analysis}

\subsection{Method}
\label{sec:lcmethod}

Estimating the peak luminosity of SNe in a given bandpass requires an
interpolation between observed data points at the time of maximum
light, followed by a $k$-correction back to some standard rest-frame
filter of interest. The interpolation step is best performed using a
smooth light curve template in the observed filter which is believed
to represent the photometric evolution of the SN; the $k$-correction
requires a time series spectral energy distribution \citep[SED;
e.g.][]{2002PASP..114..803N,2007ApJ...663.1187H}.  To parametrize the
light curve of PTF\,09dav, we fit a time series SED to the observed
photometry using the SiFTO light curve fitter
\citep{2008ApJ...681..482C} developed for SN Ia cosmology studies. We
replace the standard SN Ia spectral
template\footnote{\url{http://supernova.lbl.gov/$\sim$nugent/nugent\_templates.html}}
with one based on the sub-luminous SN Ia events SN1991bg
\citep{1992AJ....104.1543F,1993AJ....105..301L} and SN1999by
\citep{2004ApJ...613.1120G}.  Ideally one would use a SED template
which exactly matches the spectral class of the object under study.
For PTF\,09dav, the first object of its type, no such template exists,
so we use our sub-luminous template which has many common spectral
features and similar broad-band colors.

SiFTO works in flux space, manipulating a model of the SED and
synthesizing an observer-frame light curve from a given spectral
time-series in a set of filters at a given redshift, allowing an
arbitrary normalization in each observed filter (i.e., the absolute
colors of the template being fit are not important and do not
influence the fit).  The time-axis of the template is adjusted by a
dimensionless relative ``stretch'' ($s$) factor (where the input
template is defined to have $s=1$) to fit the data.  Once the
observer-frame SiFTO fit is complete, it can be used to estimate
rest-frame magnitudes in any given set of filters, provided there is
equivalent observer-frame filter coverage, and at any epoch. This is
performed by adjusting the template SED at the required epoch to have
the correct observed colors from the SIFTO fit, correcting for
extinction along the line of sight in the Milky Way, de-redshifting,
and integrating the resultant SED through the required filters. This
process is essentially a cross-filter k-correction, with the advantage
that all the observed data contribute to the SED shape used. As an
illustration, the size of the k-correction for a SN1991bg-like SN Ia
to go from observer $g$ ($r$) at $z=0.0371$ to rest-frame $B$ ($V$) is
$\sim-0.16$\,mag ($\sim-0.24$\,mag) at maximum light, and
$\sim-0.28$\,mag ($\sim-0.48$\,mag) at 10 days after maximum.

The function used to adjust the template SED can either be an
interpolating spline (in which case a perfect match in color is
attained) or, if the reason for the physical difference in the colors
between the SN and the template is understood, a function such as a
standard dust law \citep[e.g.][]{1989ApJ...345..245C} or other
wavelength-dependent color law \citep[e.g.][]{2007A&A...466...11G}.
The latter case is more useful when extrapolating beyond the
wavelength range of the observed filters, although it gives a poorer
match between observed and fit colors.

\subsection{Fit results}
\label{sec:lcfitresults}

Our SiFTO light curve fit is in Fig.~\ref{fig:lightcurve}. The fit is
good -- the reduced $\chi^2$ of the fit is 1.02 for 40 degrees of
freedom (i.e. $\chi^2=40.8/40$) -- using a standard SN Ia template in
SiFTO gives a substantially worse reduced $\chi^2$ of 4.3. Maximum
light in the rest-frame $B$-band was on 2009-08-08.9 $\pm$ 0.3 days
(or an MJD of 55051.9), i.e.  the SN was discovered some 3 days after
maximum light.  Interestingly, although the fit to the data around
maximum light is excellent, at phases beyond $\sim$+30 days there
appears to be significant excess $i$-band flux in the SN compared to
the simple prediction of the template with the $i$-band light curve
appearing to plateau, while the decay in the other filters continues
to follow the sub-luminous template. This could be due to emerging
nebular emission from the [\ion{Ca}{2}] doublet at $\lambda7291$ and
$\lambda7323$, with stronger lines at that phase than in our
SN1991bg-like sub-luminous template, and more similar to later time
spectra of calcium-rich SNe \citep{2010Natur.465..322P}.

As PTF\,09dav has no detected host galaxy at the SN position, we can
cross-check our photometric calibration using photometry synthesized
from our WHT spectrum. At the time of the WHT spectrum (+4.9 days),
SiFTO gives a light curve color of $g-R=1.17\pm0.05$, compared to a
spectral color from the WHT spectrum of $g-R=1.26$.  Given the
uncertainties in flux-calibrating long-slit spectra (e.g.
differential slit losses which will redden the SED), as well as in the
P48 $R$ and P60 $g$ filter responses used for synthetic photometry,
this is an excellent level of agreement suggesting an accurate
photometric flux calibration. A similar agreement is found for the
later P200 spectrum.

Our reported magnitudes are in the Vega system, taking our $BVR$
filter responses from \citet{1990PASP..102.1181B}, and correcting for
Milky Way extinction assuming $R_V=3.1$ and a color excess of
$E(B-V)_{\mathrm{mw}}$=0.044 \citep{1998ApJ...500..525S} with a
\citet{1989ApJ...345..245C} extinction law.  We measure peak
rest-frame absolute magnitudes at the time of $B$-band maximum light
of $M_B=-15.44\pm0.05$, $M_V=-16.01\pm0.05$ and $M_R=-16.26\pm0.05$
(errors are statistical uncertainties propagated through the light
curve fit only) using a CCM law in SiFTO to match the template to the
observed fit colors; we adopt these numbers throughout this paper.
For comparison, using a spline in SIFTO in place of the CCM law gives
$M_B=-15.54\pm0.05$, $M_V=-15.98\pm0.09$ and $M_R=-16.24\pm0.05$.

One potential concern in the light curve fit is that the P60/FTN $g$
coverage, corresponding to rest-frame $B$, did not start until
$\simeq6$ days after maximum light -- and so the $B$-band (and
$V$-band) maximum light estimate is an extrapolation (based on the
light curves synthesized from the SED template) rather than an
interpolation as with the P48 $R$-band. Our key assumption is that the
light curve of PTF\,09dav is well-represented by a ``stretched''
version of our subluminous spectral template.  While this is clearly
true in the P48 $R$ data where the fit before and after maximum light
is robust, in $g$ the data are not sufficient to test this assumption.

\subsection{Comparison to other events}
\label{sec:comp-other-events}

\begin{figure}
\plotone{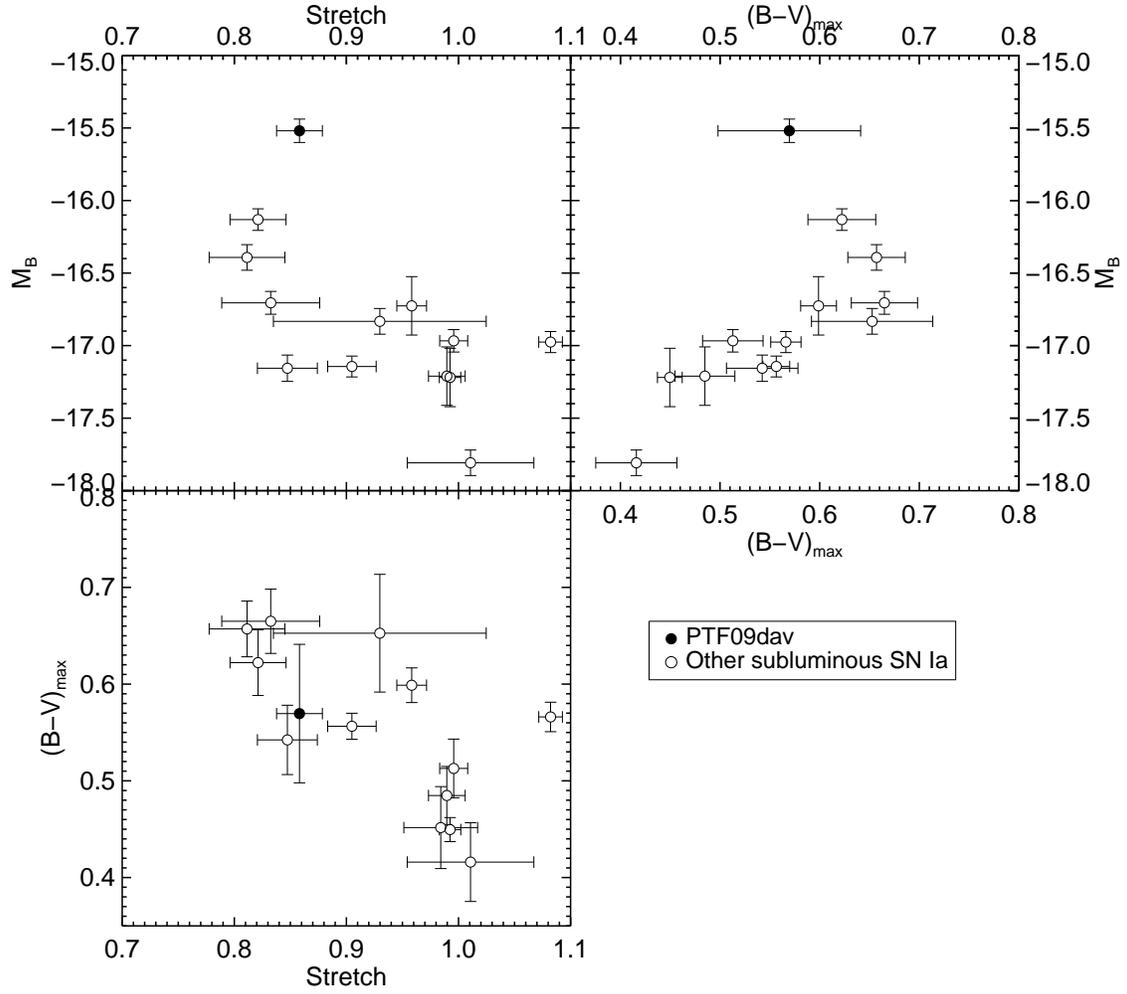}
\caption{A comparison of the photometric properties of PTF\,09dav and
  other 1991bg-like subluminous SNe Ia. Top left: stretch versus
  $M_B$, top right: $M_B$ versus color ($(B-V)_{\mathrm{max}}$), and
  lower left: stretch versus $(B-V)_{\mathrm{max}}$. Here, the stretch
  is measured relative to our sub-luminous template and is not on the
  same system as those used in cosmological studies. All magnitudes
  are corrected for Milky Way extinction. In all panels, PTF\,09dav is
  shown as the filled circle, and the other subluminous SNe Ia denoted
  by the open circles. The data are given in
  Table~\ref{tab:sublumcomp}.\label{fig:photcomp}}
\end{figure}

A comparison of PTF\,09dav to other SN1991bg-like subluminous SN Ia
events can be found in Table~\ref{tab:sublumcomp} and
Fig.~\ref{fig:photcomp}. With no commonly accepted definition for what
constitutes a SN1991bg-like event, we fit a combination of events
matching the cool classification from \citet{2009PASP..121..238B},
events with $\Delta M_{15}(B)>1.8$ from \citet{2008MNRAS.385...75T}
and \citep{2009ApJ...700..331H}, and other individual events from the
literature \citep[e.g.][]{2008ApJ...683L..29K}.  We fit all the public
SN Ia photometry with SiFTO and the same subluminous template to
ensure a consistent comparison. All magnitudes are corrected for Milky
Way extinction, and we have assumed distance moduli either calculated
from \zcmb\ when $\zcmb\ge0.01$, or taken from a redshift-independent
estimate (together with an uncertainty) from the NASA/IPAC
Extragalactic Database
(NED\footnote{\url{http://nedwww.ipac.caltech.edu/}}) in other cases.
We have also propagated through a $300\,\mathrm{km}\,\mathrm{s}^{-1}$
peculiar velocity error for those SNe Ia in the Hubble flow. We only
consider SNe with adequate photometry for a reliable light curve fit,
and those with $M_B>-18$ at maximum light, in our analysis.

The SNe show expected trends between stretch, color and $M_B$
(Fig.\ref{fig:photcomp}) which are consistent with previous studies
\citep{2004ApJ...613.1120G,2008MNRAS.385...75T,2010arXiv1011.4531G}: Fainter SNe
are redder, and with smaller values of stretch. (Note that the
stretches here are measured relative to our subluminous template and
are not on the same system as stretches of normal SNe Ia.) Clearly
PTF\,09dav is an outlier in these relations, being considerably
fainter for its color and stretch than other subluminous events:
PTF\,09dav is $\sim3.5$\,mag under-luminous compared to normal SNe Ia
($M_B\simeq-19.1$), $\sim1.5$\,mag fainter than a typical subluminous
SN Ia, and around $\sim0.5$\,mag fainter than SN2007ax, the faintest
subluminous SN Ia identified \citep{2008ApJ...683L..29K}.

No correction for host extinction (if any is present) has been applied
to the magnitudes of PTF\,09dav (or the other SNe in
Table~\ref{tab:sublumcomp}), and thus host extinction will impact
Fig.~\ref{fig:photcomp}. For PTF\,09dav,
$(B-V)_{\mathrm{max}}=0.57\pm0.08$, and when using the CCM law in
SiFTO to match the subluminous template to the fit colors for
PTF\,09dav, a small additional $E(B-V)$ of $0.08\pm0.04$ is required.
Though it may be tempting to interpret this at face value as a
reddening for the SN (and derive similar corrections for the other SNe
in Table~\ref{tab:sublumcomp}), such an approach is likely to be
dangerous. Generally the color--luminosity variation in SNe Ia is
poorly understood. There is considerable evidence for an intrinsic
color--luminosity relation which is currently impossible to
disentangle from the effects of dust unless the SN is particularly
well observed. Thus interpreting a red color as evidence of extinction
may not be correct. A second consideration is that our subluminous
template may not be appropriate for zero extinction, as it is based on
real SN photometry.  Therefore we simply note that the broad-band
colors of PTF\,09dav are similar to other sub-luminous SN Ia events and
show no evidence for significant host galaxy extinction.

We also estimate the value of the \citet{1993ApJ...413L.105P} decline
rate parameter $\dmb$ (the amount in magnitudes in the rest-frame
$B$-band that the SN declines in the 15 days following maximum light)
as $\dmb=1.87\pm0.06$. The error is a combination of propagating the
error in the stretch parameter, combined with the uncertainty in peak
$B$-band brightness. Numerous studies have fit the relation between
\dmb\ and $M_B$ for subluminous SNe Ia
\citep[e.g.,][]{2004ApJ...613.1120G,2008MNRAS.385...75T}. According to
the relation of \citet{2008MNRAS.385...75T}, our $\dmb=1.87$ should
correspond to a peak $B$ absolute magnitude of $-17.44$, nearly two
magnitudes brighter than we observe ($M_B\simeq-15.5$ corresponds to
$\dmb \simeq2.15$ according to the Taubenberger relation). This is
consistent with PTF\,09dav as an outlier in Fig.~\ref{fig:photcomp}.
However, \dmb, unlike stretch, can be misleading in these comparisons.
In the fainter examples of subluminous SNe Ia, the transition from the
fast initial decline to the slower late-time decline occurs before
$-15$ days, which distorts the relation
\citep[e.g.][]{2008ApJ...683L..29K}.

\subsection{Rise-time and energetics}
\label{sec:energetics}

The subluminous template we used in the SiFTO fit has a rise-time
($\tau_r$) from explosion of 13.0 days to $B$-band maximum light, and
14.0 days to bolometric maximum light. PTF\,09dav has a light curve
``stretch'' factor of $0.86\pm0.02$ compared to this template, giving
a rise-time to bolometric peak of $12.0\pm0.3$ days (statistical error
only).  We also estimate the rise time by fitting the P48 $R$ data in
the early rise-time region using the analytical equation
$f(t)=\alpha(\tau+\tau_r)^2$ \citep{1999AJ....118.2668R}.  Here,
$\tau=t/(1+z)$, where $t$ is the observed time, and $\alpha$ is a
normalizing constant. We find rise-times of $12.0\pm2.9$ when fitting
the first two epochs (including the first non-detection), and
$14.5\pm1.1$ when including the next epoch. These are consistent with
our estimates from the template light curve fit. The latter measure
includes data from near-maximum light, where the assumption of the
simple rise-time model may no longer hold, and should be considered a
strong upper limit.

Using our light curve fits, we can also estimate the bolometric
luminosity, \lbol, of PTF\,09dav. Estimating the bolometric luminosity
is never straightforward as full ultraviolet through infrared (or
UVOIR) light curves are rarely observed. Instead we measure the
luminosity in a given band together with the colors from our light
curve fits, and use the template to derive a bolometric luminosity
estimate (effectively a bolometric correction).  Such techniques work
well for SNe Ia events near maximum light as the peak of the
luminosity output is in the optical. Again, as with the light curve
fit, we use the template spectrum to calculate our bolometric light
curve. In principle we could use the observed spectra, but the lack of
an observation at maximum light, and the lack of $\lambda>7700$ data
in the spectrum nearest maximum, makes this difficult. Our technique
broadly matches that of \citet{2009ApJ...691..661H}.  We take the
template spectrum on the day of maximum light, adjust it using a CCM
law to have the correct color, normalize to the peak $B$-band absolute
magnitude, and integrate the resulting spectrum. The peak bolometric
luminosity (occurring $\simeq1$ day after $B$-band peak) is estimated
to be $\lbol^B=5.6\pm0.4\times 10^{41} \mathrm{erg}\,\mathrm{s}^{-1}$
(Table~\ref{tab:sublumcomp}). The implicit bolometric correction,
defined as $BC=M_{bol}-M_V$ where $M_{bol}$ is the bolometric
magnitude, is $\simeq0.2$\,mag, comparable to other sub-luminous
events SN Ia events \citep{2000A&A...359..876C}.

The ejecta mass $M_{\mathrm{ej}}$ and kinetic energy
$E_{\mathrm{kin}}$ of PTF\,09dav can be estimated from its rise time
and ejecta velocity $v_{\mathrm{ej}}$ using $M_{\mathrm{ej}}\propto
v_{\mathrm{ej}}\tau_r^2$ and $E_{\mathrm{kin}}\propto
v_{\mathrm{ej}}^3\tau_r^2$
\citep[e.g.,][]{1982ApJ...253..785A,2009AJ....138..376F}.  Comparing
PTF\,09dav to a ``typical'' SN Ia with a kinetic energy of
$10^{51}$\,ergs, a rise-time of 17.4\,days
\citep{2010ApJ...712..350H}, and a photospheric velocity of
11,000\,km\,$s^{-1}$ \citep{2005ApJ...623.1011B}, and assuming the
opacities of both are the same, we estimate an ejecta mass of
$0.36\msun$ and kinetic energy of $8\times10^{49}$\,ergs. (Here, we
have used an ejecta velocity of 6000\,km\,s$^{-1}$ for PTF\,09dav
estimated from the spectral fitting in
$\S$~\ref{sec:spectral-analysis}.) These are low values; a typical
ejecta mass for a SN Ia ranges from $\simeq0.5$\msun\ for a
SN1991bg-like event, up to the Chandrasekhar mass (1.4\msun) for
normal SNe Ia \citep[e.g.][]{2006A&A...450..241S,2007Sci...315..825M}.
A similar low ejecta mass was derived for the calcium-rich SN Ib
SN2005E \citep{2010Natur.465..322P}, and even lower ejecta masses have
been estimated for faint examples of SN2002cx-like events
\citep{2009AJ....138..376F,2010ApJ...720..704M} and for SN2002bj
\citep{2010Sci...327...58P} and SN2010X \citep{2010ApJ...723L..98K}.

If the light curve is predominantly powered by the decay of \nickel,
the \nickel\ mass, \mni, can be estimated from a knowledge of the
rise-time $\tau_r$ and \lbol\ \citep{1982ApJ...253..785A}. At maximum
light, \mni\ is given by \mni=\lbol/$\alpha$\radlum($\tau_r$)
(Arnett's rule), with \radlum\ the radioactive luminosity per solar
mass of \nickel\ from the decay to $^{56}$Co to $^{56}$Fe evaluated at
maximum light, and $\alpha$ the ratio of bolometric to radioactive
luminosities, near unity. We adopt \radlum\ as defined in
\citet{2009ApJ...691..661H} \citep[see also][]{1992ApJ...392...35B},
which includes the effect of $\tau_r$, and take $\tau_r$ and its error
from our light curve fits with a conservative additional 2 day
uncertainty in the rise time error estimate to match the range found
above.  In practice, $\alpha$ is likely to deviate from unity. For
normal SNe Ia, the light curve may peak at a luminosity that exceeds
the instantaneous radioactive luminosity ($\alpha>1$) due to a falling
temperature and opacity \citep[e.g.][]{1992ApJ...392...35B}. On the
other hand, for smaller ejecta masses some gamma rays may escape the
ejecta without being thermalized (i.e., $\alpha<1$). Using the
approach of \citet{2010ApJ...723L..98K} we estimate the optical depth
of PTF\,09dav at maximum light to be $\sim70$ indicating an effective
trapping of emitted $\gamma$-rays.  For simplicity we therefore take
$\alpha=1$, giving $\mni=0.019\pm0.003\msun$.  Under the same
assumptions, SN 2008ha produced $0.0029\msun$ of \nickel, and SN
2007ax $0.038\pm0.008\msun$ -- see Table~\ref{tab:sublumcomp} for
estimated \mni\ for all the 1991bg-like subluminous SNe. In these
calculations, we estimated $\tau_r$ for each event by multiplying the
stretch by the rise-time of the subluminous template (14 days). We
emphasize that this calculation assumes the entire light curve to be
powered by the decay of \nickel.

In summary of this section, the $M_B$, \lbol, \mni\, and ejecta mass
$M_{\mathrm{ej}}$ of PTF\,09dav are all unusually low.  PTF\,09dav is
one of the faintest subluminous SNe Ia yet discovered, and, while not
as faint as SN2008ha, it is the faintest of the SN1991bg-like
sub-class.

\section{Spectral analysis}
\label{sec:spectral-analysis}

\begin{figure}
\centering
  \includegraphics[angle=270,width=0.99\textwidth]{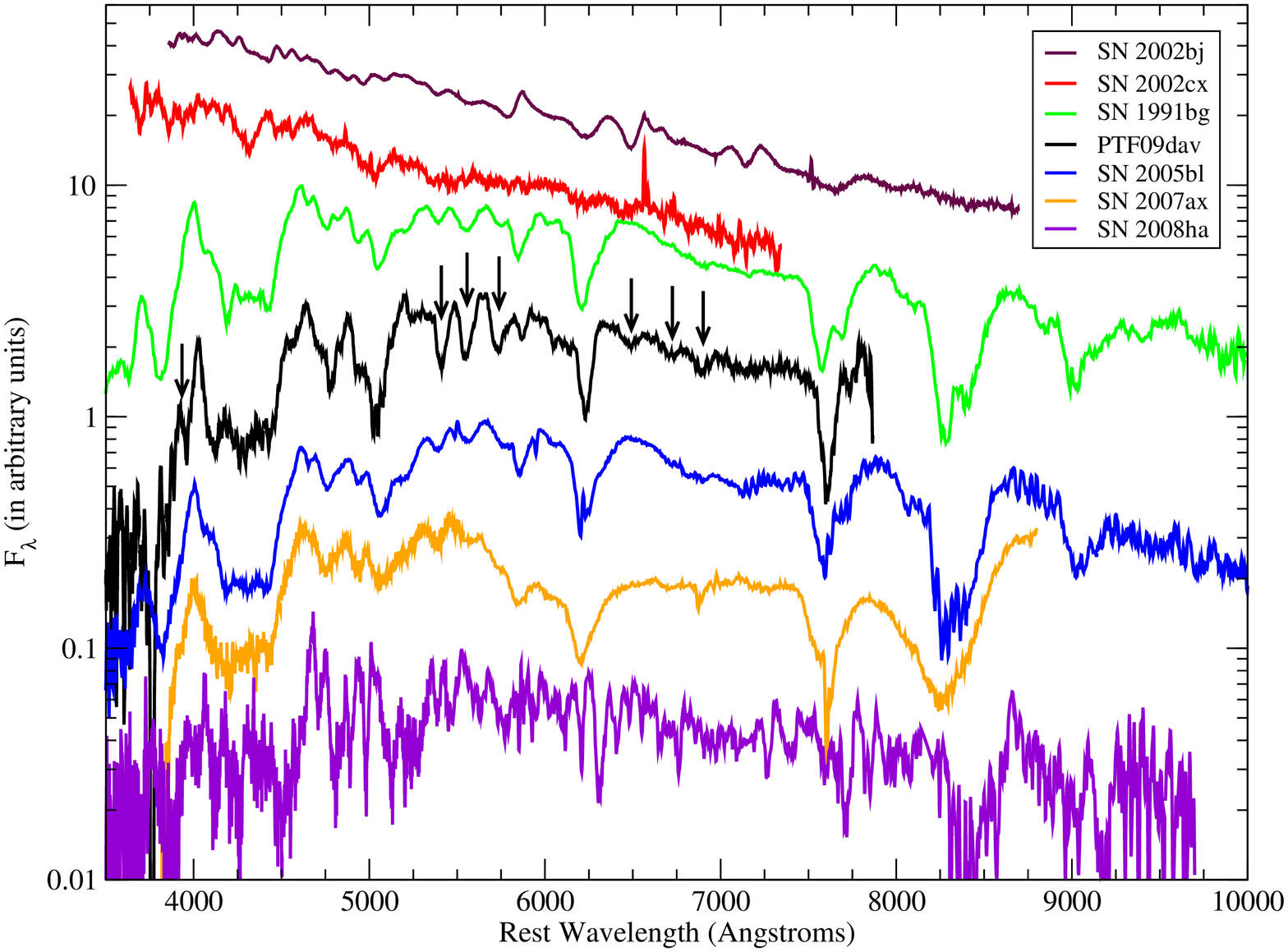}
  \caption{A comparison of the +4.9 day WHT spectrum of PTF\,09dav with
    a selection of other sub-luminous events.  From top to bottom:
    SN2002bj \citep{2010Sci...327...58P}, SN2002cx
    \citep{2003PASP..115..453L}, SN1991bg \citep{1992AJ....104.1543F},
    PTF\,09dav, SN2005bl \citep{2008MNRAS.385...75T}, SN2007ax
    \citep{2008ApJ...683L..29K}, and SN2008ha
    \citep{2009AJ....138..376F}. For each event the spectrum nearest
    to +4.9 days is shown. All spectra have been deredshifted, and
    additionally the spectra of SN1991bg and SN2005bl have been
    shifted by $-3000\,\mathrm{km}\,\mathrm{s}^{-1}$ to account for
    the velocity difference with PTF\,09dav. The vertical arrows
    highlight the principle differences between PTF\,09dav and other
    sub-luminous SNe Ia (see Fig.~\ref{fig:obsspectra2} for the
    evolution of these features in later spectra).
    \label{fig:obsspectra}}
\end{figure}

\begin{figure}
\centering
  \includegraphics[angle=270,width=0.99\textwidth]{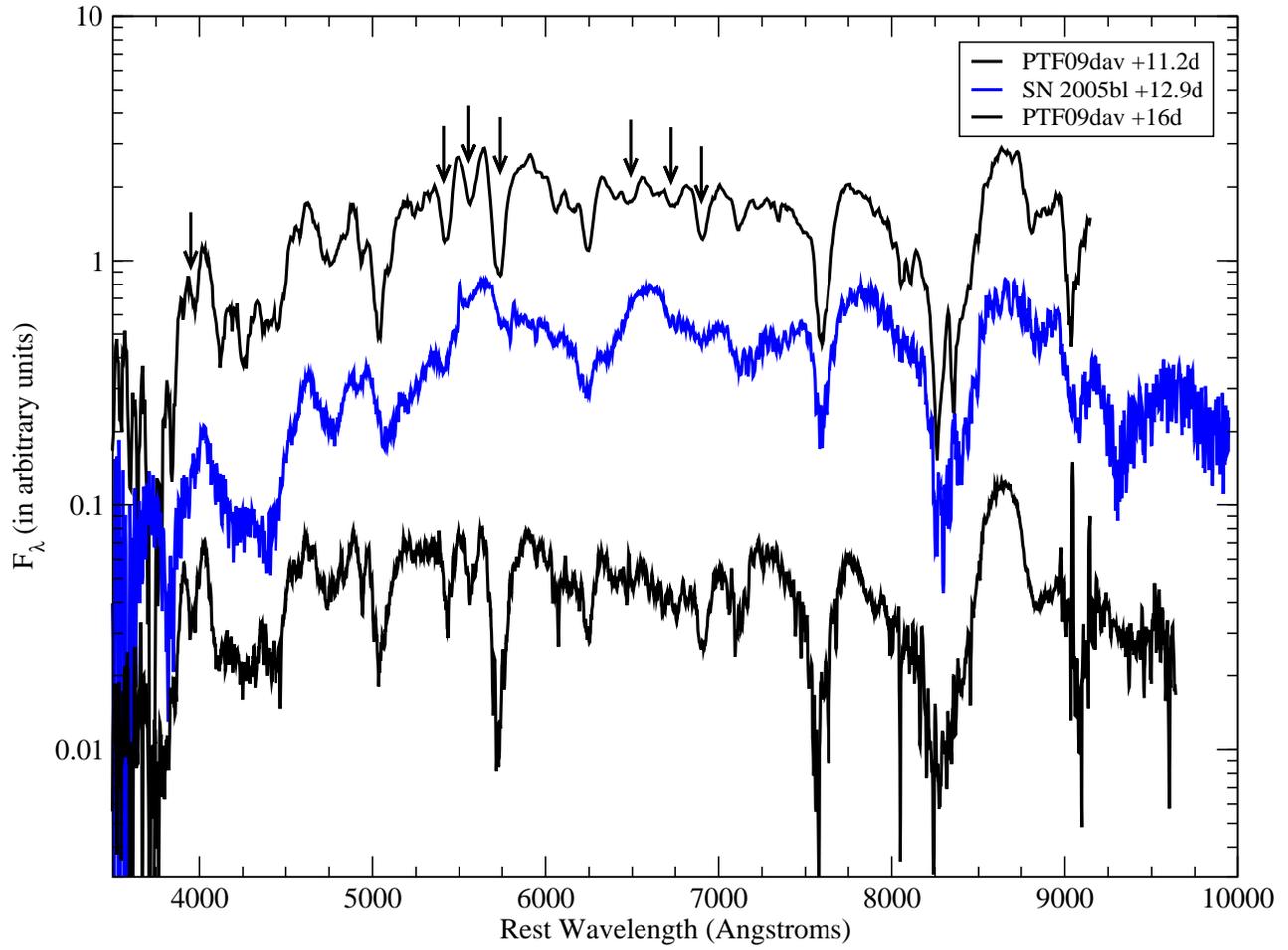}
  \caption{The +11.2d and +16d spectra of PTF\,09dav compared to
    SN2005bl at +12.9d. SN2005bl has been shifted by
    $-3000\,\mathrm{km}\,\mathrm{s}^{-1}$ to account for the velocity
    difference with PTF\,09dav. The vertical arrows mark the same
    features as in Fig.~\ref{fig:obsspectra}.
    \label{fig:obsspectra2}}
\end{figure}

We now turn to the interpretation of the PTF\,09dav spectra.  The
spectra of PTF\,09dav show many curious features compared to other
sub-luminous events (Fig.~\ref{fig:obsspectra}). The comparison to
SN2005bl \citep{2008MNRAS.385...75T} is particularly instructive, as
that spectrum was taken at an identical light curve phase (+5d).
Despite the resemblance to sub-luminous SNe Ia, with obvious
\ion{Ti}{2}, \ion{Si}{2}, and \ion{O}{1} absorption, other features in
the spectra do not appear in other sub-luminous events
(Fig.~\ref{fig:obsspectra}). In particular, absorption features at
3960\AA, 5400\AA, 5540\AA, 5720\AA, 6480\AA, 6700\AA, and 6890\AA\ do
not, at first glance, immediately correspond to features in other
sub-luminous SNe Ia. We attempt to identify these lines using an
implementation of the \synow\ code, and then discuss the results.

\subsection{Spectral fitting}
\label{sec:spectral-fitting}

When no detailed \emph{ab initio} SN explosion model provides an
immediate explanation for observations, parametrized spectrum
synthesis provides the first step towards their construction.  The
idea is to use simplified radiative transfer calculations to directly
fit SN spectra.  A good fit constrains explosion models through
interpretive spectral feature identification, with the main result
being the detection or exclusion of specific chemical elements.  The
velocity distribution of detected species within the ejecta can also
be constrained.

In our analysis we make use of \synapps\ (Thomas et al., submitted).
The physical assumptions \synapps\ uses match those of the well-known
\synow\ code \citep{2000PhDT.........6F}, so findings are restricted
to identification of features and not quantitative abundances.  But
where \synow\ is completely interactive, \synapps\ is automated.  This
relieves the user from tedious, iterative adjustment of a large number
of parameters (over 50 variables) to gain fit agreement, and assures
more exhaustive searching of the parameter space.  \synapps\ can be
thought of as the hybridization of a \synow-like calculation with a
parallel optimization framework, where spectral fit quality serves as
the objective function to optimize.

\subsection{Spectral Line Identifications}
\label{sec:spectr-line-indent}

\begin{figure}
\centering
  \includegraphics[angle=270,width=0.99\textwidth]{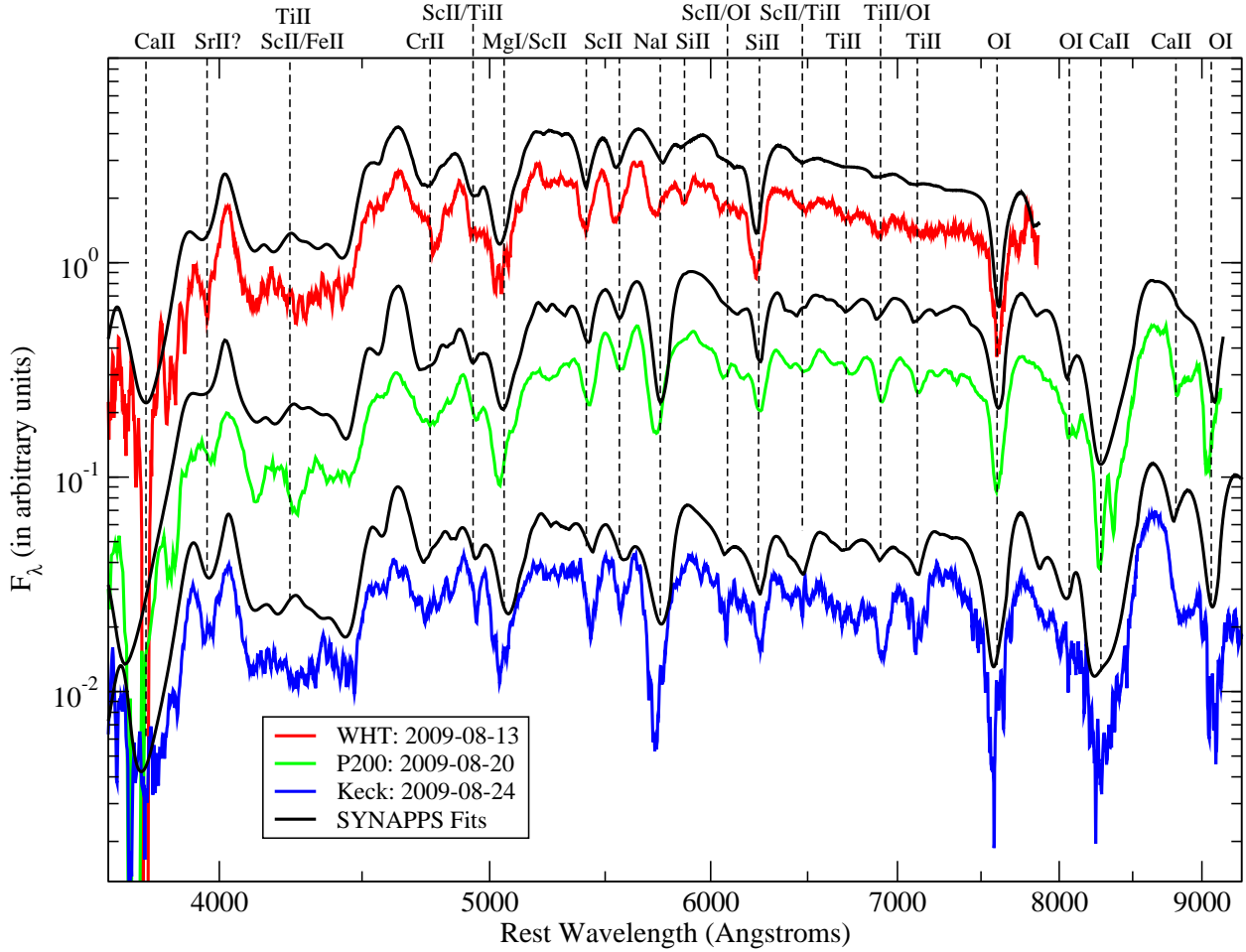}
  \caption{The three photospheric epoch spectra of PTF\,09dav from WHT
    (+4.9 days), P200 (+11.2 days) and Keck (+16.0 days), along with
    their \synapps\ fits. We have assumed a heliocentric redshift of
    0.0371. The major features in the spectrum are labeled at the
    minimum of their absorption. They include ionic species common in
    sub-luminous SNe Ia (\ion{O}{1}, \ion{Ca}{2}, \ion{Si}{2},
    \ion{Ti}{2}) together with additional lines including \ion{Sc}{2},
    and relatively strong \ion{Na}{1}.  We also tentatively identify
    \ion{Sr}{2} and \ion{Cr}{2}.  Over the period in time covered by
    these spectra, the photospheric line velocities evolve from
    $\simeq6100\,\mathrm{km}\,\mathrm{s}^{-1}$ to
    $\simeq5100\,\mathrm{km}\,\mathrm{s}^{-1}$ to
    $\simeq4600\,\mathrm{km}\,\mathrm{s}^{-1}$.\label{fig:spectralfits}}
\end{figure}

We run \synapps\ on our three photospheric spectra, and the resulting
fits are shown in Fig.~\ref{fig:spectralfits}. We identify the
following lines common to subluminous SN Ia spectra: \ion{O}{1},
\ion{Ca}{2}, \ion{Si}{2}, \ion{Ti}{2}, \ion{Fe}{2}, \ion{Co}{2}. The
\ion{Ti}{2} lines are particularly strong relative to other
SN1991bg-like events.  We also identify \ion{Sc}{2}, \ion{Na}{1}, and
\ion{Mg}{1}, as well as evidence for \ion{Sr}{2} and possibly
\ion{Cr}{2}. There was no evidence for \ion{S}{2} or \ion{Mg}{2} --
both degraded the quality of the fits, \ion{S}{2} around 5500\AA, and
\ion{Mg}{2} in the red.

\begin{figure}
\includegraphics[angle=270,width=0.495\textwidth]{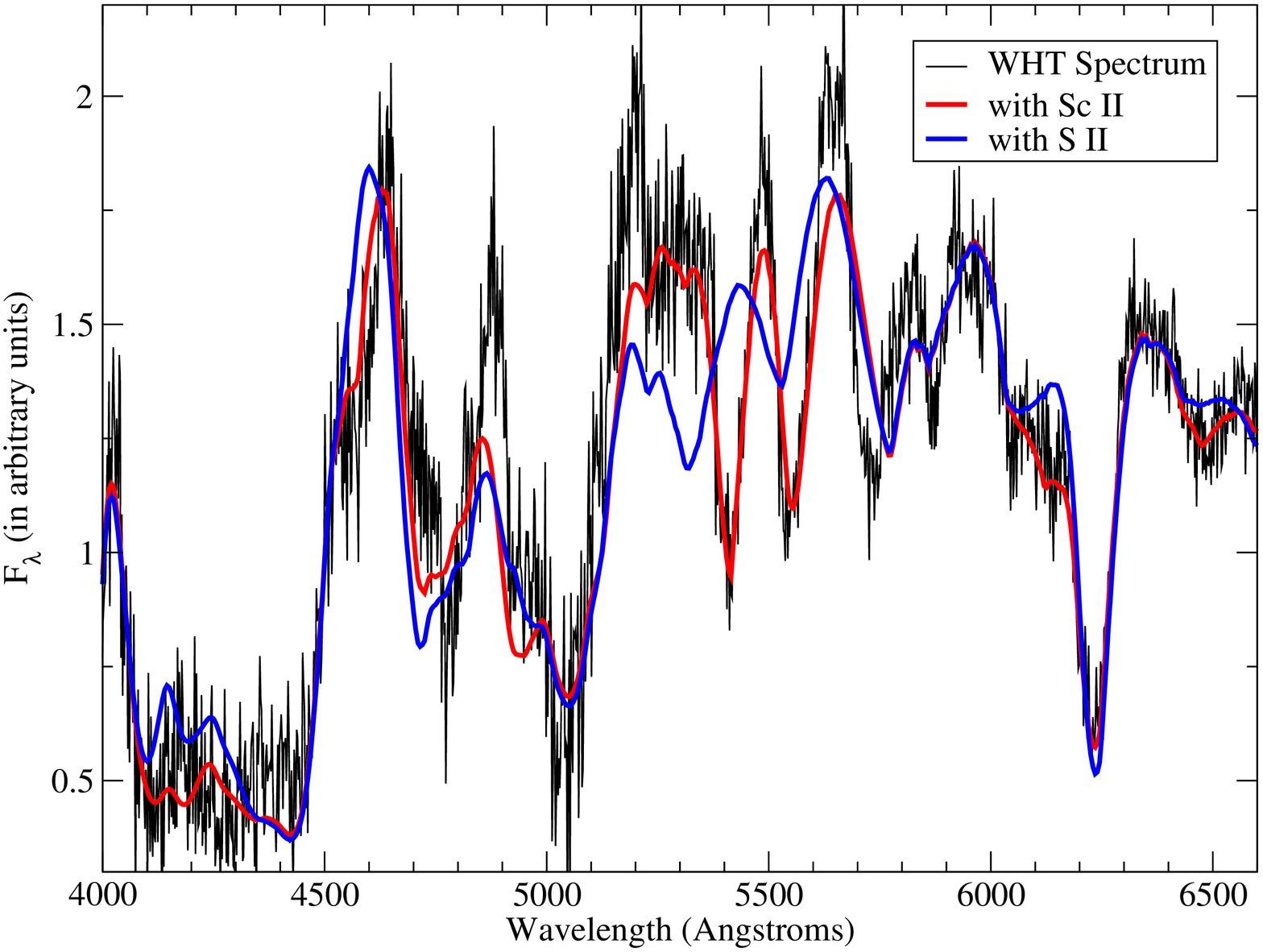}
\includegraphics[angle=270,width=0.495\textwidth]{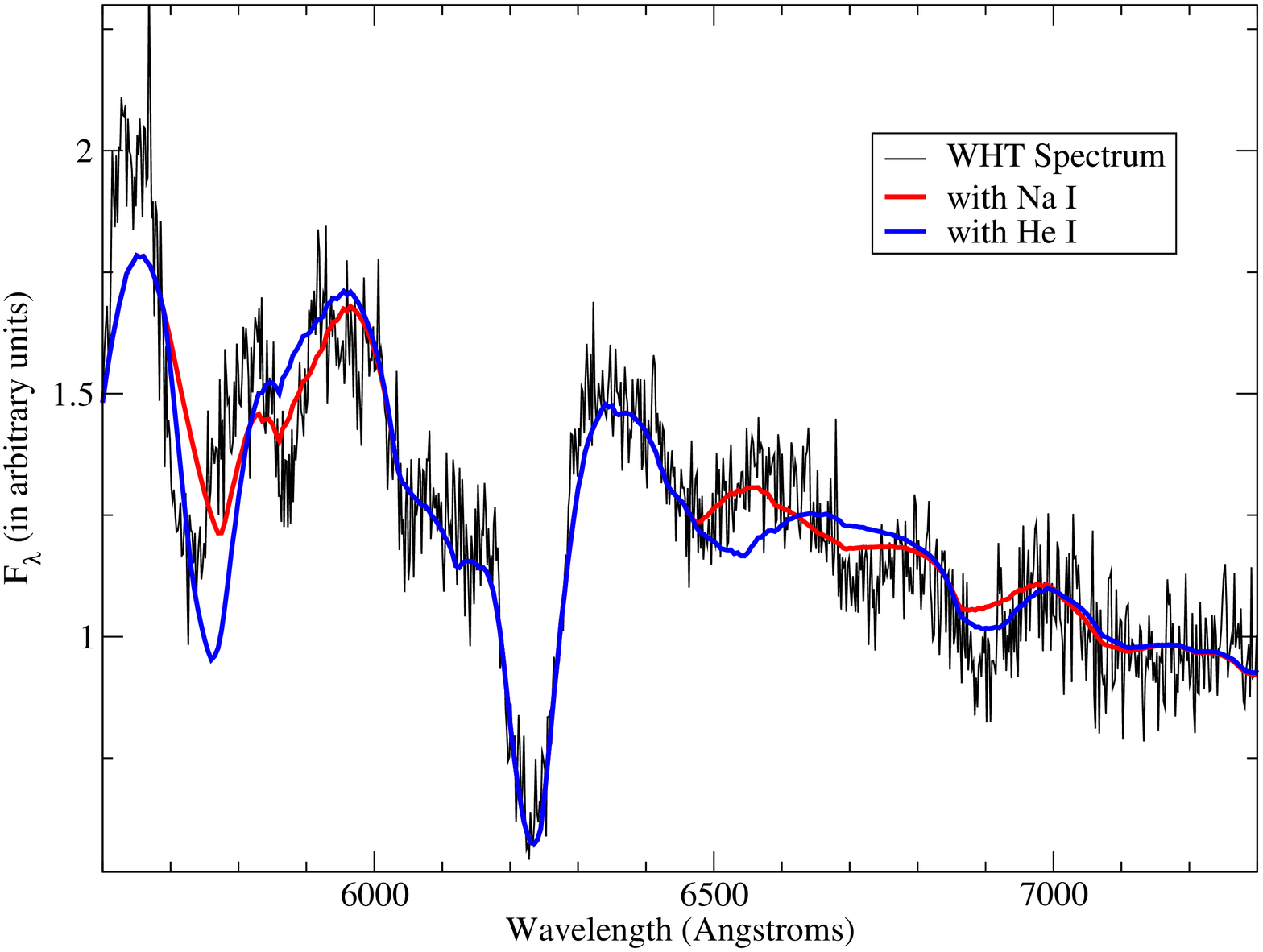}

\includegraphics[angle=270,width=0.495\textwidth]{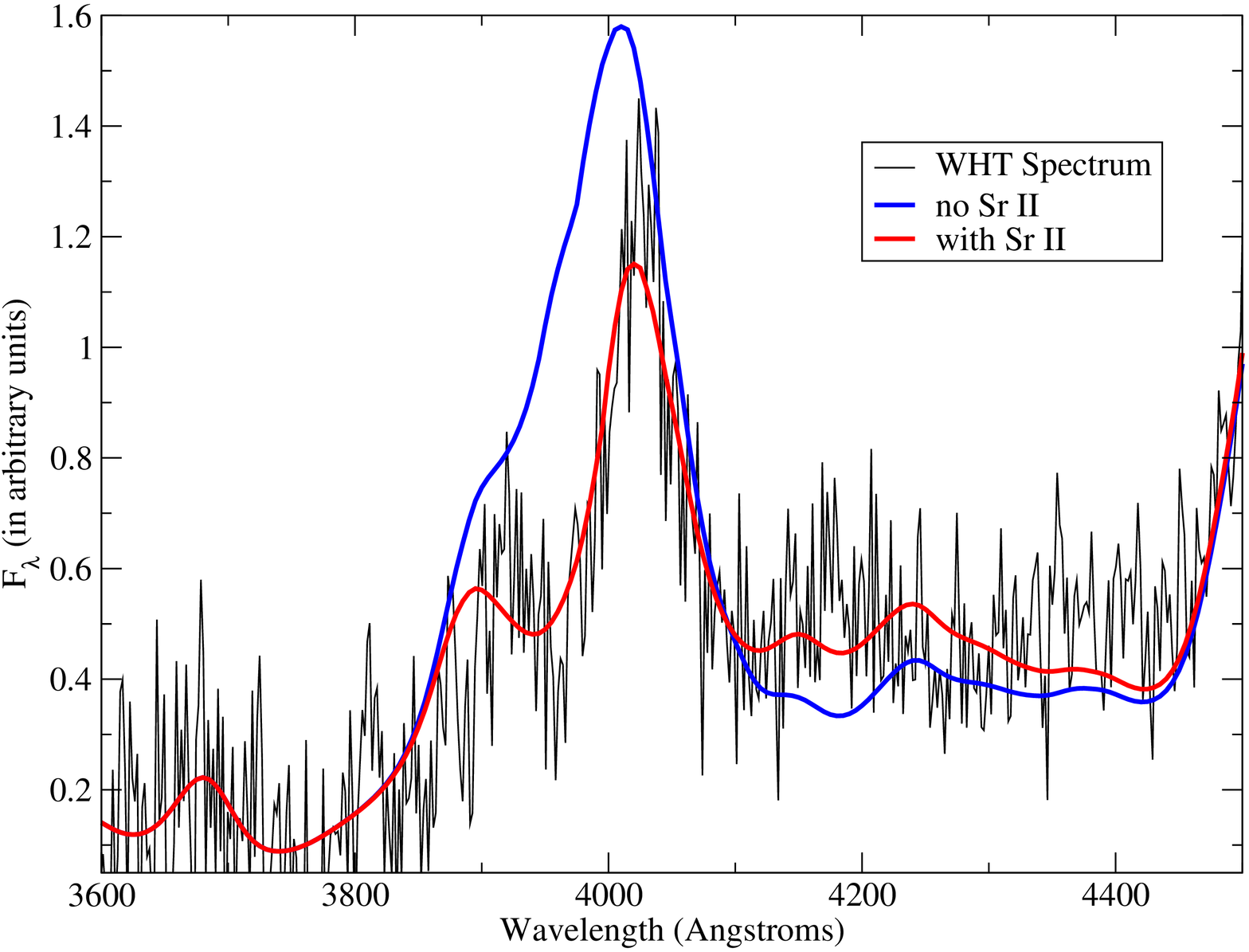}
\includegraphics[angle=270,width=0.495\textwidth]{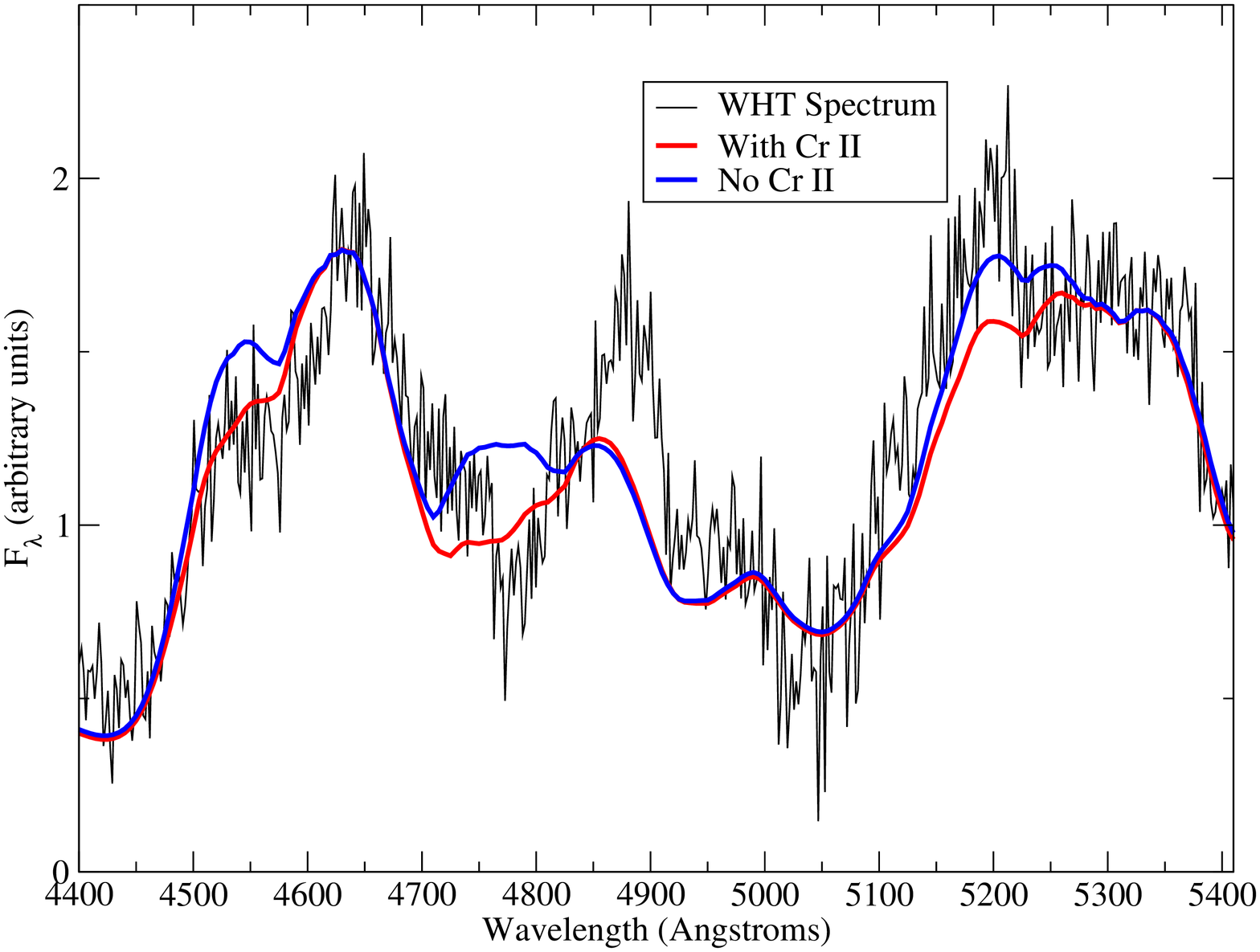}
\caption{\synapps\ fits to the +4.9 day WHT spectrum, comparing
  \ion{Sc}{2} and \ion{S}{2} (upper left), \ion{He}{1} and \ion{Na}{1}
  (upper right), and the evidence for \ion{Sr}{2} (lower left) and
  \ion{Cr}{2} (lower right).  In all panels, the black line is a
  section of the WHT spectrum, the red line our default \synapps\ fit,
  and the blue line a fit with the relevant change in the \synapps\
  set-up.  Upper left: S II clearly cannot fit features well-fit
  by Sc II, in particular the $\lambda5527$ and $\lambda5658$ lines.
  Upper right: The evidence in favor of Na or He is less clear-cut,
  although Na provides a more consistent fit. Lower left: Sr II
  improves the spectral fitting around 4000\AA.  Lower right: Cr II
  improves some regions of the fit (especially around 4800\AA), but
  degrades the fit in other areas.
  \label{fig:scs_hene}}
\end{figure}

Though unusual, the presence of \ion{Sc}{2} seems robust. As well as
the two features at 5400\AA\ and 5550\AA\ (caused by $\lambda5527$ and
$\lambda5658$), the 6490\AA\ feature (caused by $\lambda6604$),
together with improved fits between 4000--5000\AA, provides additional
confirmation. \ion{S}{2} has strong lines in this region, but cannot
be responsible for the two features at 5400\AA\ and 5550\AA\ (observer
frame) -- the ratio of the two line wavelengths does not match
\ion{S}{2}, and the velocity would be inconsistent with the other
elements (Fig.~\ref{fig:scs_hene}).  \ion{S}{2} is common in
1991bg-like events (although the lines typically become weaker and
disappear after maximum light), and its non-detection here is
therefore surprising, but it does have a higher ionization energy than
the other elements, which may point to lower temperatures.

Although \ion{Sc}{2} has been included in the \synow\ fits of faint
SN2002cx-like SNe Ia \citep{2009AJ....138..376F,2010ApJ...720..704M},
it has never been robustly detected in SNe Ia. \ion{Sc}{2} has been
observed (together with the other s-process elements \ion{Ba}{2} and
\ion{Sr}{2}) in SN1987A
\citep{1987ApJ...320L.117W,1992A&A...258..399M}, and also during the
plateau phase of some lower-luminosity SNe IIP
\citep{1998MNRAS.299..150F,2004MNRAS.347...74P}.  \ion{Mg}{1} (with no
indication of \ion{Mg}{2}) is also surprising. As both \ion{Sc}{2} and
\ion{Mg}{1} are only seen at low temperatures
\citep{1999ApJS..121..233H}, even lower than \ion{Ti}{2}, this
provides further evidence for the low temperatures of PTF\,09dav.

\ion{Na}{1} $\lambda5892$ is also highly probable
(Fig.~\ref{fig:scs_hene}), and is typically seen in subluminous events
after maximum light, although it is particularly strong in PTF\,09dav --
Fig.~\ref{fig:obsspectra2}.  This feature could also be \ion{He}{1}
$\lambda5876$ line, and we cannot unambiguously confirm \ion{Na}{1}
over \ion{He}{1} (Fig.~\ref{fig:scs_hene}).  Including \ion{He}{1} in
the fits does better match the 6900\AA\ feature (produced by
$\lambda7065$), but it introduces other features in the \synapps\
spectra which are not seen in the data, such as 6500\AA\ (caused by
$\lambda6678$). However, \synapps\ does not explicitly account for
non-local thermodynamic equilibrium effects which are known to be
important for modeling \ion{He}{1} features
\citep{1991ApJ...383..308L} -- LTE excitation in \synapps\ can
overestimate the strength of $\lambda6678$ to $\lambda5876$ and
$\lambda7065$ \citep{2003IAUS..212..346B}. Thus, all we can say is
that at least one of, and possibly both, \ion{Na}{1} and \ion{He}{1}
are present in the spectra of PTF\,09dav.

There is also reasonable evidence for the presence of \ion{Sr}{2} (via
absorption near 3960\AA), with weaker evidence for \ion{Cr}{2}
(primarily through absorption at 4770\AA), all shown in
Fig.~\ref{fig:scs_hene}.  \ion{Sr}{2} is identified through a blended
single line ($\lambda4078$ and $\lambda4216$), and has no other strong
optical features. As with \ion{Sc}{2}, \ion{Sr}{2} is also only seen
at low temperatures \citep{1999ApJS..121..233H}.  \ion{Cr}{2} is often
weakly present in SN1991bg-like SNe
\citep{1997MNRAS.284..151M,2008MNRAS.385...75T}. \ion{Ba}{2}, an
s-process element often seen with \ion{Sr}{2}, was not needed in the
fits. 

Given the apparent low temperatures, we also tested the inclusion of
\ion{S}{1} and \ion{Si}{1}, but neither was required in the fits
(their abundances came out to zero in \synapps). Indeed, \ion{S}{1}
and \ion{Si}{1} would add lines not seen in our spectra if their
abundance were non-zero. We also experimented with \ion{V}{2} and
\ion{C}{2} -- there was no evidence for either in the fits, although
they also did not degrade the fit results by adding extra lines not
seen in the spectra.

\subsection{Spectral measurements}
\label{sec:spectral-measurements}

We measure ejecta velocities from the blueshift of the absorption
minima of the P-Cygni line profiles. This process can carry a number
of uncertainties, particularly the blending of neighboring lines, so
we use the \synapps\ results which effectively average over several
lines. The line velocities evolve from
$\simeq6100\,\mathrm{km}\,\mathrm{s}^{-1}$ (+4.9 days) to
$\simeq5100\,\mathrm{km}\,\mathrm{s}^{-1}$ (+11.2 days) to
$\simeq4600\,\mathrm{km}\,\mathrm{s}^{-1}$ (+16.0 days), and the major
lines all share a common velocity.

The strength of the \ion{Na}{1} (or \ion{He}{1}) line increases with
time, and our tentative \ion{Sr}{2} detection also becomes stronger.
\ion{Ti}{2}, already stronger than in most SN1991bg-like events, also
increases in strength with phase. By contrast, the evidence for
\ion{Fe}{2}, as well as \ion{Cr}{2}, has disappeared by +16 days;
those lines are not needed in the later-epoch \synapps\ fits.

We also estimate the \citet{1995ApJ...455L.147N} \ion{Si}{2} ratio,
$\mathcal{R}(\mathrm{Si})$, defined as the ratio of the depth of the
\ion{Si}{2} features at $\lambda5972$ and $\lambda6355$. We calculate
this ratio as 0.35, smaller then the values typical of sub-luminous
SNe Ia of $\sim0.65$ \citep{2008MNRAS.385...75T}, and naively
indicating a higher temperature than in those events. However, the
presence of strong emission from the \ion{Na}{1} (or \ion{He}{1}) line
blue-wards of the $\lambda5972$ feature will significantly affect this
measurement, decreasing the depth of the $\lambda5972$ feature and
hence artificially decreasing the observed ratio. We also measure the
(pseudo) equivalent widths of these two \ion{Si}{2} features, which
can be used to differentiate sub-types of SNe Ia using the
W(6100)/W(5750) plane \citep{2006PASP..118..560B,2009PASP..121..238B}.
Using the notation of \citet{2006PASP..118..560B}, W(6100) (the
$\lambda6355$ line) is 45\AA\ and W(5750) (the $\lambda5972$ line) is
10\AA. These measurements are both significantly lower than in
SN1991bg-like events (typical values of $\simeq$100\AA\ and
$\simeq$45\AA\ respectively), and therefore lie well outside the
region in W(6100)/W(5750) space containing other cool SNe with
significant \ion{Ti}{2} absorption \citep{2009PASP..121..238B}.

\section{Discussion}
\label{sec:discussion}

PTF\,09dav is an unusual supernova. Though superficially similar to a
subluminous SN Ia, it is considerably fainter at $M_B\simeq-15.5$
(without being redder), and is one of the fastest SNe Ia observed. The
peculiarity extends to its spectra, which have low ejecta velocities
($\sim6000\,\mathrm{km}\,\mathrm{s}^{-1}$) and unusual spectral lines,
including an unambiguous detection of \ion{Sc}{2} -- which we believe
to be unique to PTF\,09dav among SNe Ia. The origin of Sc is not clear.
Though it is the decay product of $^{44}$Ti, the half-life is $\sim60$
years, which makes that an unlikely source for the PTF\,09dav maximum
light spectra. $^{45}$Ti, with a half-life of $\sim3$ hours, or higher
and more unstable Ti isotopes, are more plausible if they are
synthesized in large enough quantities. Some $^{45}$Ti is predicted to
be synthesized in the delayed detonation SN Ia explosion models of
\citet{2010ApJ...712..624M}, but only with very small mass fractions.
The low temperatures also probably contribute to the enhanced
visibility of Sc. If real, the \ion{Sr}{2} detection is also unusual,
and presumably originates from the s-process in the progenitor star.

PTF\,09dav does share some similarities with SN2005E and the sub-class
of calcium-rich Type Ib/c SNe \citep{2010Natur.465..322P}. SN2005E was
faint ($M_B=-14.8$), as are all calcium-rich SNe, with a similar low
ejecta mass to PTF\,09dav (but higher ejecta velocities). SN2005E had
a Ca rich spectrum with no prominent intermediate-mass C/O-burning
products (e.g., Mg, S, Si), and although PTF\,09dav is also
Ca/Ti-rich, unlike SN2005E it does show the presence of Si and Mg
lines.  Perhaps more interestingly, the environments are very similar.
The environment of PTF\,09dav is discrepant compared to other
subluminous SNe Ia: PTF\,09dav is either lying on the extreme
outskirts of a spiral galaxy ($\sim41$kpc from the center, projected
distance), in contrast to the massive E/S0 galaxies that form the bulk
of the SN1991bg-like host population, or is located in a very faint,
undetected dwarf host galaxy ($M>-12.8$).  Such a host galaxy would be
one of the faintest on record; the SN1999aa-like SN Ia SN1999aw had a
host absolute magnitude of $M_V=-12.4$ \citep{2002AJ....124.2905S},
but this was a slow-declining luminous event, in contrast to
PTF\,09dav.  Similarly, SN2005E was also located some distance
($\simeq23$kpc projected) from the center of its host galaxy, outside
the plane of the disc, with no evidence for a faint dwarf host (or
indeed any star formation activity) at the SN position.

The lack of a detection of a faint host galaxy at the position of
PTF\,09dav argues against a core-collapse origin and points to an older
progenitor. The likelihood of finding a high-mass star far out in the
galactic halo of a massive galaxy is discussed in detail by
\citet{2010Natur.465..322P}, who conclude that a low-mass, old stellar
progenitor is substantially more plausible at these large radii.  A
high-velocity and high-mass star ejected from the disk or center of
the host of PTF\,09dav is always conceivable, although an $8\msun$ star
would have to be ejected from the center at
$\sim1000\,\mathrm{km}\,\mathrm{s}^{-1}$ to reach a distance of
41\,kpc in its $\sim40$\,Myr lifetime. While such stars have been
observed in our galaxy
\citep{2005ApJ...634L.181E,2008A&A...483L..21H}, they are rare, and
represent only a tiny fraction of massive stars in the Milky Way
\citep{2010Natur.465..322P}. They would also, by themselves, provide
no explanation for the odd properties of PTF\,09dav -- the chances of
PTF detecting a SN from such a star and it simultaneously having
peculiar (unique) properties, seem low. Of course, the future
detection of a faint host for PTF\,09dav would certainly weight the
argument in favor of a core-collapse origin -- deeper imaging is
underway (Kasliwal et al., in prep.).

At such a large galactic radius, the most likely progenitor type for
PTF\,09dav is an old (possibly metal-poor) white dwarf star, perhaps
even located in a globular cluster
\citep[e.g.][]{2009ApJ...695L.111P}. One simple physical explanation
for PTF\,09dav is that it is just a fainter example of a normal
SN1991bg-like SN. The enhanced \ion{Ti}{2} compared to brighter
SN1991bg-like events, together with \ion{Sc}{2} and \ion{Mg}{1} and a
lack of \ion{S}{2}, point to lower temperatures consistent with the
small inferred \mni, and the light curve is well represented by a
stretched version of SN1991bg.  However, the SN also appears quite
calcium-rich, particularly at later phases, with nebular [\ion{Ca}{2}]
$\lambda7307$ beginning to emerge in the +16day spectrum, and an
$i$-band excess consistent with strong [\ion{Ca}{2}] emission at day
+30 and beyond.  This scenario also provides no immediate explanation
for strong \ion{Sc}{2} in the spectra (other than the cool
temperatures that would at least allow any \ion{Sc}{2} present to be
seen), and, if the possible \ion{He}{1} in the spectrum is real and
not just Na, then PTF\,09dav is a very different event to other
SN1991bg-like SNe.

Since calcium-rich spectra could arise from the detonation of He,
\citet{2010Natur.465..322P} propose that SN2005E-like SNe be explained
by the detonation of helium on a CO core white dwarf
\citep[e.g.][]{1982ApJ...253..798N,1982ApJ...257..780N,1986ApJ...301..601W,1994ApJ...423..371W,1995ApJ...452...62L}.
These models are also worth exploring in the context of PTF\,09dav.
Various contemporary models of the observable outcomes of these
sub-Chandrasekhar-mass explosions exist, with a variety of CO cores,
ignition densities, and He shell masses modeled, and which consider
both full star (CO core plus shell) and shell-only explosions
\citep{2007ApJ...662L..95B,2010ApJ...715..767S,2010ApJ...714L..52S,2010A&A...514A..53F,2010arXiv1009.3829W,2010arXiv1010.5292W}.
The `full-star' explosions tend to be too luminous to explain the
properties of sub-luminous SNe, but the helium shell detonation models
\citep[][]{2010ApJ...715..767S,2010arXiv1009.3829W} are more
promising.

The light curves tend to be powered by the decay of \nickel,
$^{48}$Cr, $^{52}$Fe plus other radioactive nuclei and their decay
products, but the key predicted observables of such scenarios can vary
significantly according to the configuration of the progenitor system.
\citet{2010ApJ...715..767S} consider CO cores of $>0.6$\msun, and
their models are typified by faint ($M_R=-15$ to $-18$), rapidly
rising ($<$10 days) and possibly multi-peaked light curves which
redden quickly following maximum light due to the emissivity of
Fe-group isotopes when they transition from the doubly to
singly-ionized state \citep[e.g.][]{2006ApJ...649..939K}. The spectra
are dominated by \ion{Ca}{2} and \ion{Ti}{2} features at high velocity
($\sim10^4\,\mathrm{km}\,\mathrm{s}^{-1}$), lacking intermediate mass
elements (such as silicon). Lower CO core masses of $<0.6$\msun\ have
been explored by \citet{2010arXiv1009.3829W}, who show that at these
low masses the abundance of \nickel\ drops sharply, with light curves
powered predominantly by $^{48}$Cr, with the abundances of Ca and Ti
both increasing together with the amount of unburnt helium. The
presence of unburnt helium in the observed spectra of all He
detonation models is unclear due to the LTE assumptions in the
\citet{2010ApJ...715..767S} and \citet{2010arXiv1009.3829W} models,
but some He might be expected.

The observations of PTF\,09dav match several of these features. The
spectra contain significant \ion{Ca}{2} and \ion{Ti}{2}. The light
curve also peaks with a consistent luminosity, and becomes red in
(e.g.) $r-i$ at later phases, although this may also be plausibly
explained by the emergence of the nebular [\ion{Ca}{2}] doublet at
$\lambda7307$.  The ejecta velocity and rise-time are both on the edge
of the model outcomes predicted by \citet{2010ApJ...715..767S}, although
more consistent with the lower CO-core mass models of
\citet{2010arXiv1009.3829W}.  However, there is no immediate
explanation for the presence of \ion{Sc}{2} (which is not present in
any of the model spectra), and the spectra of PTF\,09dav also contain
both \ion{Si}{2} and \ion{O}{1}, which are predicted to have very low
abundances in these helium detonation events. Thus PTF\,09dav is
unlikely to be the result of a pure helium detonation, at least in the
forms currently considered in the literature.

Another possibility for PTF\,09dav is the deflagration of a helium
shell layer on the surface of a CO white dwarf, which may occur for
smaller He shell masses \citep{2010arXiv1010.5292W}.  These can have
quite different properties to those of helium shell detonations, with
the dim light curve ($M_B\sim-15$) again powered mostly by $^{48}$Cr
(rather than $^{56}$Ni) leading to abundant Ti in the spectra. As the
burning is incomplete, as with the detonation models some helium
should also remain. The explosion energy per unit ejected mass is
smaller than for the detonation models, giving slower light curves
than for detonation models of the same brightness, with relatively low
ejecta velocities $\sim4000\,\mathrm{km}\,\mathrm{s}^{-1}$.  The
presence of Sc in the spectra (from $^{45}$Ti) is a prediction in the
helium deflagration models \citep{2010arXiv1010.5292W}, although it is
dependent on the neutron excess and hence metallicity. However, these
deflagration models also predict quite blue optical colors ($B-V\sim
0$) with peak luminosities fainter than PTF\,09dav, and, as for the
detonation models, there is a lack of Si/Mg in the model spectra. In
this regard, PTF\,09dav seems to combine both C/O-burning and
He-burning products -- further detailed modeling and comparisons with
observations will be required for a full understanding.

Whatever the physical explanation, PTF\,09dav illustrates the discovery
power of new high-cadence transient surveys such as PTF. Although the
`faint and fast' observational parameter space remains relatively
unexplored, these early new discoveries are able to test our
understanding of cosmic explosions and motivate new directions in
explosion modeling.

\acknowledgements 

We acknowledge useful discussions with Brian Schmidt and Dan Kasen. We
thank Richard Ellis for providing the P200/DBSP spectrum. MS
acknowledges support from the Royal Society. EOO is supported by an
Einstein fellowship and NASA grants. LB is supported by the National
Science Foundation under grants PHY 05-51164 and AST 07-07633. SBC
acknowledges generous support from Gary and Cynthia Bengier and the
Richard and Rhoda Goldman Foundation.  DP is supported by an Einstein
fellowship.  The WHT is operated on the island of La Palma by the
Isaac Newton Group in the Spanish Observatorio del Roque de los
Muchachos of the Instituto de Astrofísica de Canarias.  Observations
obtained with the Samuel Oschin Telescope and the 60-inch Telescope at
the Palomar Observatory as part of the Palomar Transient Factory
project, a scientific collaboration between the California Institute
of Technology, Columbia University, Las Cumbres Observatory, the
Lawrence Berkeley National Laboratory, the National Energy Research
Scientific Computing Center, the University of Oxford, and the
Weizmann Institute of Science.  Some of the data presented herein were
obtained at the W.M. Keck Observatory, which is operated as a
scientific partnership among the California Institute of Technology,
the University of California and the National Aeronautics and Space
Administration. The Observatory was made possible by the generous
financial support of the W.M. Keck Foundation. This research has made
use of the NASA/IPAC Extragalactic Database (NED) which is operated by
the Jet Propulsion Laboratory, California Institute of Technology,
under contract with the National Aeronautics and Space Administration.
The Weizmann Institute participation of in PTF is supported in part by
grants from the Israeli Science Foundation to AG. Weizmann-Caltech
collaborative work on PTF is supported by a grant from the Binational
Science Foundation (BSF) to AG and SRK. Collaborative work by AG and
MS is supported by a grant from the Weizmann-UK ``making connection''
program. Collaborative work of AG and PAM is supported by a
Weizmann-Minerva grant. AG is supported by an EU/FP7 Marie Curie IRG
fellowship and a research grant from the Peter and Patricia Gruber
Awards.

{\it Facilities:} \facility{PO:1.2m}, \facility{ING:Herschel}, \facility{PO:1.5m}, \facility{Hale}, \facility{Keck:I}, \facility{LCOGT}

\clearpage
\begin{deluxetable}{lccccccccl}
\tablecaption{Photometry for PTF\,09dav. Photometry is listed for a common zeropoint of 30\tablenotemark{a}.}
\tablehead{\colhead{MJD\tablenotemark{b}} & \multicolumn{2}{c}{$g$} & \multicolumn{2}{c}{$r$} & \multicolumn{2}{c}{$R$} & \multicolumn{2}{c}{$i$} & \colhead{Telescope}\\
\colhead{} & \colhead{counts} &\colhead{error} & \colhead{counts} &\colhead{error} & \colhead{counts} &\colhead{error} & \colhead{counts} &\colhead{error} & \colhead{}}
\startdata
55039.32 & \nodata        & \nodata       & \nodata              & \nodata  & \phs\phn\phn 278.5    & \phn 557.0 & \nodata & \nodata & P48 \\
55045.21 & \nodata        & \nodata       & \nodata              & \nodata  & \phs\phn 3476.9       & 1430.7     & \nodata & \nodata & P48 \\
55045.31 & \nodata        & \nodata       & \nodata              & \nodata  & \phs\phn 3300.0       & \phn 726.0 & \nodata & \nodata & P48 \\
55049.43 & \nodata        & \nodata       & \nodata              & \nodata  & \phs\phn 8683.0       & 1621.0     & \nodata & \nodata & P48 \\
55050.20 & \nodata        & \nodata       & \nodata              & \nodata  & \phs 14791.5          & 4045.6     & \nodata & \nodata & P48 \\
55050.39 & \nodata        & \nodata       & \nodata              & \nodata  & \phs\phn 9281.0       & 1799.0     & \nodata & \nodata & P48 \\
55054.29 & \nodata        & \nodata       & \nodata              & \nodata  & \phs 14438.4          & 1234.6     & \nodata & \nodata & P48 \\
55054.40 & \nodata        & \nodata       & \nodata              & \nodata  & \phs 11930.8          & \phn 865.1 & \nodata & \nodata & P48 \\
55058.29 & \phs 3292.3    & \phn 244.1    & \nodata              & \nodata  & \nodata               & \nodata    & \nodata & \nodata & P60 \\
55058.42 & \nodata        & \nodata       & \phs 10508.5         &    346.9 & \nodata               & \nodata    & \nodata & \nodata & P60 \\
55058.42 & \nodata        & \nodata       & \nodata              & \nodata  & \nodata               & \nodata    & 13501.6 &    601.9 & P60 \\
55059.41 & \nodata        & \nodata       & \nodata              & \nodata  & \phs\phn 9980.0       & \phn 403.0 & \nodata & \nodata & P48 \\
55060.23 & \nodata        & \nodata       & \nodata              & \nodata  & \phs\phn 8964.8       & \phn 594.3 & \nodata & \nodata & P48 \\
55060.25 & \nodata        & \nodata       & \nodata              & \nodata  & \nodata               & \nodata    & 14045.7 &    681.9 & P60 \\
55061.26 & \nodata        & \nodata       & \nodata              & \nodata  & \nodata               & \nodata    & 12955.2 &    512.7 & P60 \\
55061.37 & \phs 1887.6    & \phn 109.0    & \nodata              & \nodata  & \nodata               & \nodata    & \nodata & \nodata & FTN \\
55061.37 & \nodata        & \nodata       & \nodata              & \nodata  & \nodata               & \nodata    & 11874.5 &    254.2 & FTN \\
55062.34 & \nodata        & \nodata       & \nodata              & \nodata  & \phs\phn 7782.0       & \phn 471.0 & \nodata & \nodata & P48 \\
55062.40 & \nodata        & \nodata       & \nodata              & \nodata  & \phs\phn 7353.0       & \phn 542.0 & \nodata & \nodata & P48 \\
55064.36 & \phs 1594.8    & \phn 107.0    & \nodata              & \nodata  & \nodata               & \nodata    & \nodata & \nodata & P60 \\
55064.36 & \nodata        & \nodata       & \phs\phn 5706.7      &    217.3 & \nodata               & \nodata    & \nodata & \nodata & P60 \\
55066.39 & \nodata        & \nodata       & \nodata              & \nodata  & \phs\phn 4843.1       & \phn 514.2 & \nodata & \nodata & P48 \\
55066.43 & \nodata        & \nodata       & \nodata              & \nodata  & \nodata               & \nodata    & \phn 8154.8 &    423.1 & P60 \\
55067.17 & \nodata        & \nodata       & \nodata              & \nodata  & \phs\phn 4906.7       & \phn 551.5 & \nodata & \nodata & P48 \\
55067.45 & \nodata        & \nodata       & \nodata              & \nodata  & \nodata               & \nodata    & \phn 7173.5 &    429.3 & P60 \\
55068.45 & \phs 1154.2    & \phn 133.2    & \nodata              & \nodata  & \nodata               & \nodata    & \nodata & \nodata & P60 \\
55068.45 & \nodata        & \nodata       & \phs\phn 4449.1      &    206.4 & \nodata               & \nodata    & \nodata & \nodata & P60 \\
55068.46 & \nodata        & \nodata       & \nodata              & \nodata  & \nodata               & \nodata    & \phn 4945.3 &    480.3 & P60 \\
55069.35 & \phs 1068.0    & \phn 122.9    & \nodata              & \nodata  & \nodata               & \nodata    & \nodata & \nodata & FTN \\
55069.35 & \nodata        & \nodata       & \phs\phn 3964.3      &    156.1 & \nodata               & \nodata    & \nodata & \nodata & FTN \\
55069.35 & \nodata        & \nodata       & \nodata              & \nodata  & \nodata               & \nodata    & \phn 6374.3 &    314.6 & FTN \\
55069.42 & \phs 1172.7    & \phn\phn 72.6 & \nodata              & \nodata  & \nodata               & \nodata    & \nodata & \nodata & FTN \\
55069.43 & \nodata        & \nodata       & \phs\phn 3766.2      &    110.0 & \nodata               & \nodata    & \nodata & \nodata & FTN \\
55069.43 & \nodata        & \nodata       & \nodata              & \nodata  & \nodata               & \nodata    & \phn 6898.8 &    209.0 & FTN \\
55069.46 & \phs 1115.6    & \phn 117.5    & \nodata              & \nodata  & \nodata               & \nodata    & \nodata & \nodata & P60 \\
55069.46 & \nodata        & \nodata       & \phs\phn 4243.1      &    191.0 & \nodata               & \nodata    & \nodata & \nodata & P60 \\
55069.46 & \nodata        & \nodata       & \nodata              & \nodata  & \nodata               & \nodata    & \phn 6702.6 &    292.5 & P60 \\
55069.59 & \phs\phn 939.6 & \phn\phn 93.0 & \nodata              & \nodata  & \nodata               & \nodata    & \nodata & \nodata & FTN \\
55069.59 & \nodata        & \nodata       & \phs\phn 3893.7      &    138.7 & \nodata               & \nodata    & \nodata & \nodata & FTN \\
55069.60 & \nodata        & \nodata       & \nodata              & \nodata  & \nodata               & \nodata    & \phn 6591.3 &    220.9 & FTN \\
55071.54 & \phs 1052.9    & \phn 119.4    & \nodata              & \nodata  & \nodata               & \nodata    & \nodata & \nodata & FTN \\
55071.55 & \nodata        & \nodata       & \phs\phn 3450.9      &    178.2 & \nodata               & \nodata    & \nodata & \nodata & FTN \\
55071.55 & \nodata        & \nodata       & \nodata              & \nodata  & \nodata               & \nodata    & \phn 5585.9 &    248.2 & FTN \\
55072.46 & \phs\phn 901.7 & \phn 82.6     & \nodata              & \nodata  & \nodata               & \nodata    & \nodata & \nodata & FTN \\
55072.46 & \nodata        & \nodata       & \phs\phn 3307.5      &    169.4 & \nodata               & \nodata    & \nodata & \nodata & FTN \\
55072.47 & \nodata        & \nodata       & \nodata              & \nodata  & \nodata               & \nodata    & \phn 5377.1 &    171.4 & FTN \\
55074.45 & \phn $-$201.6  & \phn 289.7    & \nodata              & \nodata  & \nodata               & \nodata    & \nodata & \nodata & FTN \\
55074.46 & \nodata        & \nodata       & \phs\phn 2868.6      &    252.2 & \nodata               & \nodata    & \nodata & \nodata & FTN \\
55074.46 & \nodata        & \nodata       & \nodata              & \nodata  & \nodata               & \nodata    & \phn 4928.4 &    359.2 & FTN \\
55074.60 & \nodata        & \nodata       & \phs\phn 2974.2      &    378.5 & \nodata               & \nodata    & \nodata & \nodata & FTN \\
55074.61 & \nodata        & \nodata       & \nodata              & \nodata  & \nodata               & \nodata    & \phn 4887.1 &    536.0 & FTN \\
55076.59 & \phs 1089.1    & \phn 509.4    & \nodata              & \nodata  & \nodata               & \nodata    & \nodata & \nodata & FTN \\
55076.60 & \nodata        & \nodata       & \phs\phn 3494.6      &    591.3 & \nodata               & \nodata    & \nodata & \nodata & FTN \\
55083.54 & \nodata        & \nodata       & \phs\phn 1811.6      &    364.3 & \nodata               & \nodata    & \nodata & \nodata & FTN \\
55087.31 & \nodata        & \nodata       & \nodata              & \nodata  & \phs\phn\phn\phn 56.0 & \phn 494.0 & \nodata & \nodata & P48 \\
55087.47 & \phs\phn 526.7 & \phn 131.9    & \nodata              & \nodata  & \nodata               & \nodata    & \nodata & \nodata & FTN \\
55087.47 & \nodata        & \nodata       & \phs\phn 1167.9      &    196.1 & \nodata               & \nodata    & \nodata & \nodata & FTN \\
55087.48 & \nodata        & \nodata       & \nodata              & \nodata  & \nodata               & \nodata    & \phn 3470.4 &    206.9 & FTN \\
55088.16 & \nodata        & \nodata       & \nodata              & \nodata  & \phs\phn 1406.8       & \phn 477.0 & \nodata & \nodata & P48 \\
55088.21 & \nodata        & \nodata       & \nodata              & \nodata  & \phs\phn 1563.0       & \phn 339.0 & \nodata & \nodata & P48 \\
55093.50 & \phs\phn 205.2 & \phn 168.6    & \nodata              & \nodata  & \nodata               & \nodata    & \nodata & \nodata & FTN \\
55093.51 & \nodata        & \nodata       & \phs\phn\phn 457.8   &    203.6 & \nodata               & \nodata    & \nodata & \nodata & FTN \\
55093.51 & \nodata        & \nodata       & \nodata              & \nodata  & \nodata               & \nodata    & \phn 2900.7 &    336.5 & FTN \\
55095.42 & \nodata        & \nodata       & \phs\phn\phn 921.9   &    308.9 & \nodata               & \nodata    & \nodata & \nodata & P60 \\
55095.42 & \nodata        & \nodata       & \nodata              & \nodata  & \nodata               & \nodata    & \phn 2579.6 &    363.5 & P60 \\
55098.39 & \phs\phn 339.2 & \phn\phn 70.1 & \nodata              & \nodata  & \nodata               & \nodata    & \nodata & \nodata & FTN \\
55098.39 & \nodata        & \nodata       & \phn\phn\phn $-$66.3 &    128.0 & \nodata               & \nodata    & \nodata & \nodata & FTN \\
55098.40 & \nodata        & \nodata       & \nodata              & \nodata  & \nodata               & \nodata    & \phn 2260.3 &    206.4 & FTN \\
55102.42 & \nodata        & \nodata       & \phs\phn\phn 763.3   &    238.0 & \nodata               & \nodata    & \nodata & \nodata & FTN \\
55102.42 & \nodata        & \nodata       & \nodata              & \nodata  & \nodata               & \nodata    & \phn 2757.1 &    268.0 & FTN \\
55123.25 & \nodata        & \nodata       & \nodata              & \nodata  & \phs\phn 2179.4       & \phn 584.9 & \nodata & \nodata & P48 \\
55138.20 & \nodata        & \nodata       & \nodata              & \nodata  & \phn\phn $-$659.2     & 1475.4     & \nodata & \nodata & P48 \\
55142.19 & \nodata        & \nodata       & \nodata              & \nodata  & \phs\phn\phn 272.3    & \phn 704.7 & \nodata & \nodata & P48 \\
\enddata
\tablenotetext{a}{Convert (positive) counts, $c$, to Vega magnitudes, $m$,
  using $m=-2.5\log(c)+30$}
\tablenotetext{b}{Modified Julian Date; $\mathrm{MJD}=\mathrm{JD}-2400000.5$ }
\label{tab:photometry}
\end{deluxetable}

\begin{deluxetable}{cccccc}
  \tablecaption{Photometric properties of PTF\,09dav compared to
    1991bg-like  SNe Ia} \tablehead{\colhead{Name} &
    \colhead{$M_B$\tablenotemark{a}} &
    \colhead{$(B-V)_{\mathrm{max}}$} &
    \colhead{``stretch''\tablenotemark{b}} & \colhead{\lbol\tablenotemark{c}($\times
      10^{42}\,\mathrm{erg}\,\mathrm{s}^{-1}$)} & \colhead{\mni\tablenotemark{d}
      (\msun)}} \tablecolumns{6} \startdata
PTF09dav & -15.44$\pm$0.08 & 0.56$\pm$0.07 & 0.86$\pm$0.02 & 0.56$\pm$0.04 & 0.019$\pm$0.003\\
sn2007ax & -16.15$\pm$0.17 & 0.62$\pm$0.03 & 0.82$\pm$0.02 & 1.13$\pm$0.18 & 0.038$\pm$0.008\\
sn2007al & -16.41$\pm$0.17 & 0.65$\pm$0.03 & 0.81$\pm$0.03 & 1.50$\pm$0.23 & 0.050$\pm$0.010\\
sn1991bg & -16.71$\pm$0.20 & 0.66$\pm$0.03 & 0.83$\pm$0.04 & 2.01$\pm$0.37 & 0.068$\pm$0.016\\
sn2005bl & -16.72$\pm$0.09 & 0.60$\pm$0.02 & 0.96$\pm$0.01 & 1.89$\pm$0.15 & 0.072$\pm$0.011\\
sn2006je & -16.85$\pm$0.07 & 0.64$\pm$0.06 & 0.93$\pm$0.09 & 2.24$\pm$0.15 & 0.083$\pm$0.014\\
sn1998de & -16.97$\pm$0.07 & 0.51$\pm$0.03 & 1.00$\pm$0.01 & 2.13$\pm$0.15 & 0.084$\pm$0.012\\
sn2005ke & -16.98$\pm$0.20 & 0.57$\pm$0.02 & 1.08$\pm$0.01 & 2.29$\pm$0.42 & 0.096$\pm$0.021\\
sn1999da & -17.03$\pm$0.02 & 0.48$\pm$0.03 & 0.99$\pm$0.02 & 2.18$\pm$0.05 & 0.085$\pm$0.011\\
sn2006bz & -17.16$\pm$0.08 & 0.55$\pm$0.01 & 0.90$\pm$0.02 & 2.65$\pm$0.20 & 0.096$\pm$0.014\\
sn1997cn & -17.16$\pm$0.13 & 0.54$\pm$0.04 & 0.85$\pm$0.03 & 2.61$\pm$0.32 & 0.090$\pm$0.016\\
sn1999da & -17.21$\pm$0.09 & 0.48$\pm$0.03 & 0.99$\pm$0.02 & 2.59$\pm$0.22 & 0.101$\pm$0.015\\
sn1999by & -17.22$\pm$0.20 & 0.44$\pm$0.01 & 0.99$\pm$0.01 & 2.52$\pm$0.47 & 0.098$\pm$0.022\\
sn1998bp & -17.81$\pm$0.11 & 0.41$\pm$0.04 & 1.01$\pm$0.06 & 4.18$\pm$0.43 & 0.166$\pm$0.028\\
sn2007au & -18.20$\pm$0.03 & 0.20$\pm$0.04 & 1.16$\pm$0.04 & 5.01$\pm$0.13 & 0.224$\pm$0.027\\
\enddata
\tablenotetext{a}{Corrected for Milky Way extinction but not any host galaxy extinction. Assumes $H_0=70$\,km\,s$^{-1}$\,Mpc$^{-1}$}
\tablenotetext{b}{Relative to our subluminous SN Ia template. We emphasize that these stretches cannot be compared to stretches derived using a normal SN Ia light curve template which have a very different definition of $s=1$.}
\tablenotetext{c}{Estimated at the peak in the bolometric light curve}
\tablenotetext{d}{Assumes that $\tau_r$ for each SN is given by the SN stretch in column 4 multiplied by the template rise-time of 14 days.}
\label{tab:sublumcomp}
\end{deluxetable}

\end{document}